\documentclass[iop]{emulateapj}
\usepackage{natbib}

\def\kepler{{\slshape Kepler}}
\def\spitzer{{\it Spitzer}}

\begin{document}

\title{Kepler-93b: A Terrestrial World Measured to Within 120 km, and a Test Case for a New Spitzer Observing Mode}

\author{Sarah~Ballard\altaffilmark{1,2}, 
William~J.~Chaplin\altaffilmark{3,4},
David~Charbonneau\altaffilmark{5}, 
Jean-Michel~D\'esert\altaffilmark{6}, 
Francois Fressin\altaffilmark{5}, 
Li~Zeng\altaffilmark{5}, 
Michael~W.~Werner \altaffilmark{7} 
   Guy~R.~Davies\altaffilmark{3,4},
   Victor~Silva Aguirre\altaffilmark{4},
   Sarbani~Basu\altaffilmark{8},
   J{\o}rgen~Christensen-Dalsgaard\altaffilmark{4},
   Travis~S.~Metcalfe\altaffilmark{9,4},
   Dennis~Stello\altaffilmark{10},
   Timothy~R.~Bedding\altaffilmark{10},
   Tiago~L.~Campante\altaffilmark{3,4},
   Rasmus~Handberg\altaffilmark{3,4},
   Christoffer~Karoff\altaffilmark{4},
   Yvonne~Elsworth\altaffilmark{3,4},
   Ronald~L.~Gilliland\altaffilmark{11},
   Saskia~Hekker\altaffilmark{12,13,3},
   Daniel~Huber\altaffilmark{14,15}, 
   Steven~D.~Kawaler\altaffilmark{16},
   Hans~Kjeldsen\altaffilmark{4},
   Mikkel~N.~Lund\altaffilmark{4},
   Mia~Lundkvist\altaffilmark{4}
}

\altaffiltext{1}{University of Washington, Seattle, WA 98195, USA; sarahba@uw.edu}
\altaffiltext{2}{NASA Carl Sagan Fellow}
\altaffiltext{3}{School of Physics and Astronomy, University of
  Birmingham, Edgbaston, Birmingham, B15 2TT, UK}
\altaffiltext{4}{Stellar Astrophysics Centre (SAC), Department of
  Physics and Astronomy, Aarhus University, Ny Munkegade 120, DK-8000
  Aarhus C, Denmark}
\altaffiltext{5}{Harvard-Smithsonian Center for Astrophysics, Cambridge, MA 02138}
\altaffiltext{6}{Department of Astrophysical and Planetary Sciences, University of Colorado, Boulder CO 80309}
\altaffiltext{7}{Jet Propulsion Laboratory, California Institute of
  Technology, Pasadena CA 91125 USA}
\altaffiltext{8}{Department of Astronomy, Yale University, New Haven,
  CT, 06520, USA}
\altaffiltext{9}{Space Science Institute, Boulder, CO 80301, USA}
\altaffiltext{10}{Sydney Institute for Astronomy, School of Physics,
  University of Sydney 2006, Australia}
\altaffiltext{11}{Center for Exoplanets and Habitable Worlds, The
  Pennsylvania State University, University Park, PA, 16802, USA}
\altaffiltext{12}{Max-Planck-Institut f\"ur Sonnensystemforschung, Justus-von-Liebig-Weg 3, 37077 G\"ottingen, Germany}
\altaffiltext{13}{Astronomical Institute, ``Anton Pannekoek'',
  University of Amsterdam, The Netherlands}
\altaffiltext{14}{NASA Ames Research Center, Moffett Field,
  CA 94035, USA}
\altaffiltext{15}{SETI Institute, 189 Bernardo Avenue, Mountain View, CA 94043, USA } 
\altaffiltext{16}{Department of Physics and Astronomy, Iowa State
  University, Ames, IA, 50011, USA}

\keywords{eclipses  ---  stars: planetary systems --- stars: individual (KOI 69, KIC 3544595)}

\begin{abstract}
We present the characterization of the Kepler-93 exoplanetary system, based on three years of photometry gathered by the \kepler\ spacecraft. The duration and cadence of the \kepler\ observations, in tandem with the brightness of the star, enable unusually precise constraints on both the planet and its host. We conduct an asteroseismic analysis of the \kepler\ photometry and conclude that the star has an average density of 1.652$\pm$0.006 g cm$^{-3}$. Its mass of 0.911$\pm$0.033 $M_{\odot}$ renders it one of the lowest-mass subjects of asteroseismic study. An analysis of the transit signature produced by the planet Kepler-93b, which appears with a period of 4.72673978$\pm$9.7$\times$10$^{-7}$ days, returns a consistent but less precise measurement of the stellar density, 1.72$^{+0.02}_{-0.28}$ g cm$^{-3}$. The agreement of these two values lends credence to the planetary interpretation of the transit signal. The achromatic transit depth, as compared between \kepler\ and the \spitzer\ Space Telescope, supports the same conclusion. We observed seven transits of Kepler-93b with \spitzer, three of which we conducted in a new observing mode. The pointing strategy we employed to gather this subset of observations halved our uncertainty on the transit radius ratio $R_{P}/R_{\star}$. 
We find, after folding together the stellar radius measurement of 0.919$\pm$0.011 $R_{\odot}$ with the transit depth, a best-fit value for the planetary radius of 1.481$\pm$0.019 $R_{\oplus}$. The uncertainty of 120 km on our measurement of the planet's size currently renders it one of the most precisely measured planetary radii outside of the Solar System. Together with the radius, the planetary mass of 3.8$\pm$1.5 $M_{\oplus}$ corresponds to a rocky density of 6.3$\pm$2.6 g cm$^{-3}$. After applying a prior on the plausible maximum densities of similarly-sized worlds between 1--1.5 $R_{\oplus}$, we find that Kepler-93b possesses an average density within this group.
\end{abstract}

\section{Introduction}
The number of confirmed exoplanets now stands at 1,750 \citep{Lissauer14, Rowe14}. This figure excludes the additional thousands of candidates identified by the Kepler spacecraft, which possess a mean false positive rate of 10\% \citep{Morton11, Fressin13}. The wealth of data from NASA's \kepler\ mission in particular has enabled statistical results on the ubiquity of small exoplanets. \cite{Fressin13} investigated planet occurrence for stellar spectral types ranging from F to M,  while \cite{Petigura13a,Petigura13b} focused upon sunlike stars and both \cite{Dressing13} and \cite{Swift13} on smaller M dwarfs . However, of the more than 4000 planetary candidates identified by \kepler, only 58 possess measured masses\footnote[1]{From exoplanets.org, accessed on 17 March 2014}. We define a mass measurement to be a detection of the planet with 95\% confidence, from either radial velocity measurements or transit timing variations. The time-intensive nature of follow-up observations renders only a small subsample amenable to such detailed study. Kepler-93, a very bright ($V$ magnitude of 10) host star to a rocky planet, is a member of this singular group. It has been the subject of a dedicated campaign of observations spanning the 3 years since its identification as a candidate \citep{Borucki11}. 

%We have plumbed the \kepler\ photometry for signatures of the stellar oscillations, we have measured the line-of-sight velocity to the star over a period of three years, and we have observed the planetary transit with the \spitzer\ Space Telescope. 

An asteroseismic investigation underpins our understanding of the host star. Asteroseismology, which employs the oscillation timescales of stars from their brightness or velocity variations, is a powerful means of probing stellar interiors. Asteroseismic measurements derived from photometry require a long observational baseline at high cadence (to detect typically mHz frequencies) and photometric precision at the level of 10 ppm. Where these measurements are achievable, they constrain stellar densities with a precision of 1\% and stellar ages within 10\% \citep{Brown94, Silva13}. The technique can enrich the studies of transiting exoplanets, whose photometric signature independently probes the stellar density. The ratio of the semi-major axis to the radius of the host star ($a/R_{\star}$) is a transit observable property, and is directly related to the mean density of the host star (\citealt{Seager03}, \citealt{Sozzetti07} and \citealt{Torres08}). \cite{Nutzman11} were the first to apply an asteroseismic density measurement of an exoplanet host star, in tandem with a transit light curve, to refine knowledge of the planet. With transit photometry gathered with the Fine Guidance Sensor on the {\it Hubble} Space Telescope, they employed the asteroseismic solution derived by \cite{Gilliland11} to constrain their transit fit. This method resulted in a three-fold precision improvement on the radius of HD 17156b (1.0870$\pm$0.0066 $R_{J}$ compared to 1.095$\pm$ 0.020 $R_{J}$). The \kepler\ mission's long baselines and unprecedented photometric precision make asteroseismic studies of exoplanet hosts possible on large scales. \cite{Batalha11} used asteroseimic priors on the host star in their study of Kepler's first rocky planet Kepler-10b, recently updated by \cite{Fogtmann14}. Asteroseismic characterization of the host star also informed the exoplanetary studies of \cite{Howell12}, \cite{Borucki12}, \cite{Carter12}, \cite{Barclay13}, \cite{Chaplin13}, and \cite{Gilliland13}.  \cite{Huber13} increased the number of \kepler\ exoplanet host stars with asteroseismic solutions to 77.  Kepler-93 is a rare example of a sub-solar mass main-sequence dwarf that is bright enough to yield high-quality data for asteroseismology. Intrinsically faint, cool dwarfs show weaker-amplitude oscillations than their more luminous cousins. These targets are scientifically valuable not only as exoplanet hosts, but also as test beds for stellar interior physics in the sub-solar mass regime. 

In addition to its science merit as a rocky planet host, the brightness of Kepler-93 made it an optimal test subject for a new observing mode with the \spitzer\ Space Telescope. \spitzer\, and in particular its Infrared Array Camera (IRAC, \citealt{Fazio04}), has a rich history of enhancing exoplanetary studies. Previous studies with IRAC include maps of planetary weather \citep{Knutson07}, characterization of super-Earth atmospheres \citep{Desert11}, and the detection of new worlds \citep{Demory11b}. Applications of post-cryogenic \spitzer\ to the \kepler\ planets address in largest part the false-positive hypothesis. An authentic planet will present the same transit depth, independent of the wavelength at which we observe it. Combined with other pieces of evidence, Warm \spitzer\ observations at 4.5 $\mu$m contributed to the validations of a number of Kepler Objects of Interest. These planets include Kepler-10c \citep{Fressin11}, Kepler-14b \citep{Buchave11}, Kepler-18c \& d \citep{Cochran11}, Kepler-19b \citep{Ballard11b}, Kepler-22b \citep{Borucki12}, Kepler-25b \& c \citep{Steffen12}, Kepler-20c, d, e, \& f \citep{Gautier12, Fressin12}, Kepler-61b \citep{Ballard13}, and Kepler-401b \citep{Vaneylen14}. The growing list of transiting exoplanets includes ever smaller planets around dimmer stars. \spitzer\ observations of their transits are more challenging as the size of the astrophysical signal approaches the size of instrumental sources of noise. The \spitzer\ Science Center developed the ``peak-up" observational technique,  detailed in \cite{Ingalls12} and \cite{Grillmair12}, to improve the ultimate precision achievable with IRAC. Kepler-93 had been the subject of \spitzer\ observations prior to these improvements to test the false-positive hypothesis for the system. These pre-existing \spitzer\ observations of the star, combined with the intrinsic brightness of the target (which enables \spitzer\ to peak-up on the science target rather than an adjacent star within an acceptance magnitude range), made Kepler-93b an ideal test subject for the efficacy of peak-up. To this end, \spitzer\ observed a total of 7 full transits of Kepler-93b on the same pixel, 4 without peak-up mode and 3 with peak-up mode enabled. This data set allows us to investigate the effectiveness of the peak-up pointing strategy for \spitzer.

In Section 2 we present our analysis of the \kepler\ transit light curve. We describe our measurement and characterization of the asteroseismic spectrum of the star, as well as the transit
parameters. In Section 3, we describe our reduction and analysis of seven \spitzer\ light curves of Kepler-93, and detail the effects of the
peak-up observing mode to the photometry. We go on in Section 4 to present evidence for the authentic planetary nature of Kepler-93b. Our reasoning is based on the consistency of the asteroseismic density and the density inferred from the transit light curve, the transit depth observed by the \spitzer\ Space Telescope, the radial velocity observations of the star, and other evidence. In Section 5, we comment on possible planetary compositions. We conclude in Section 6.

\section{{\slshape Kepler} Observations and Analysis}
\cite{Argabright08} provided an overview of the
{\slshape Kepler} instrument, and \cite{Caldwell10} and
\cite{Jenkins10} provided a summary of its performance since
launch. \cite{Borucki11} first identified Kepler Object of Interest (KOI) 69.01 as an exoplanetary candidate (\kepler\ Input Catalog number
3544595). Our analysis of Kepler-93 is based upon 37 months of
short cadence data collected in \emph{Kepler} observing quarters Q2.3
through Q14.3, inclusive. \emph{Kepler} short-cadence
(SC) \citep{Gilliland10} sampling, characterized by a 58.5 s
exposure time, is crucial to the detection of the short-period asteroseismic oscillations presented by solar-type stars (see also \citealt{Chaplin11a}). Kepler-93 was a subject of the asteroseismic short-cadence survey, conductexd during the first 10 months of \emph{Kepler} operations \citep{Chaplin11a}. They gathered one month of data per target. Only the brightest among the sub-solar mass stars possessed high enough signal-to-noise data to yield a detection in such a short time (e.g., see \citealt{Chaplin11b}). With the \emph{Kepler} instrument, they continued to gather dedicated, long-term observations of the best of these targets, Kepler-93 among them. We employed the light curves generated by the \kepler\ aperture photometry (PDC-Map) pipeline, described by \cite{Smith12} and \cite{Stumpe12}.

% In addition, short cadence observations of Kepler Objects of Interest (KOIs) is an equally important source of low-mass asteroseismic targets \citep{Huber13}, with observing durations extended further to yield detections.

\subsection{Asteroseismic Estimation of Fundamental Stellar Properties}
\label{sec:prop}

\begin{figure*}
\epsscale{0.8}
\plotone{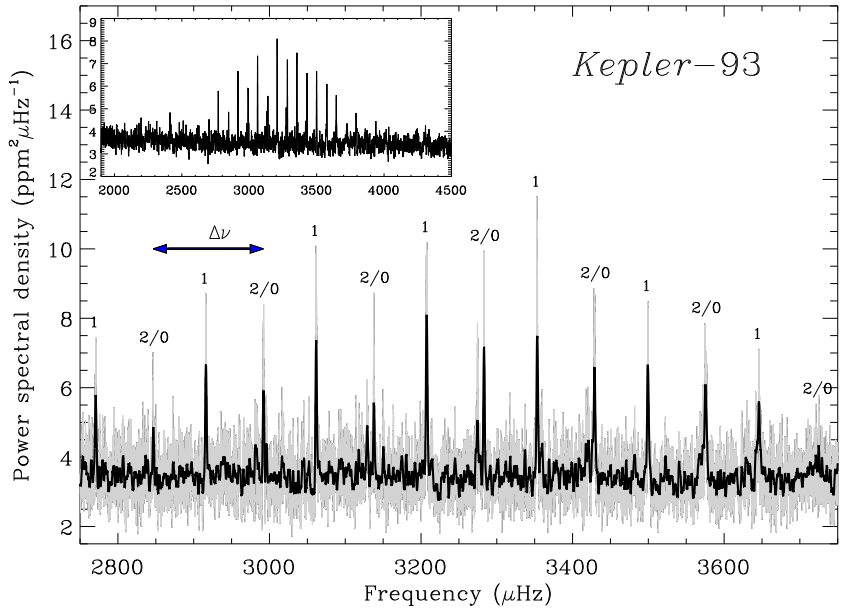}

\caption{Power spectrum of Kepler-93. The main plot shows a close-up of
  the strongest oscillation modes, tagged according to their angular
  degree, $l$. The large frequency separation, here between a pair of
  adjacent $l=0$ modes, is also marked. The black and grey curves show
  the power spectrum after smoothing with boxcars of widths 1.5 and
  $0.4\,\rm \mu Hz$, respectively. The inset shows the full extent of the observable
  oscillations.}

\label{fig:powspec}
\end{figure*}

To prepare the \kepler\ photometry for asteroseismic study, we first removed planetary transits from the light curve by applying a median high-pass filter (Handberg et al., in preparation). We then performed the asteroseismic analysis on the Fourier power spectrum of the filtered light curve. Fig.~\ref{fig:powspec} shows the resulting power spectrum of Kepler-93. We find a typical spectrum of solar-like oscillations, with several overtones of
acoustic (pressure, or p) modes of high radial order, $n$, clearly detectable in the data. Solar-type stars oscillate in both radial and non-radial modes, with frequencies $\nu_{nl}$, with $l$ the angular degree. Kepler-93 shows detectable overtones of modes with $l \le 2$. The dominant frequency spacing is the large separation $\Delta\nu_{nl} =
\nu_{n+1\,l} - \nu_{n\,l}$ between consecutive overtones $n$ of the same degree $l$.

%%%%%%%%%%%%%%%%%%%%%%%%%%%%%%%%%%%%%%%%%%%%%%%%%%%%%%%%%%%%%%%%%%%%%%%

%%%%%%%%%%%%%%%%%%%%%%%%%%%%%%%%%%%%%%%%%%%%%%%%%%%%%%%%%%%%%%%%%%%%%%%

\cite{Huber13} reported asteroseismic properties based on the use of only
average or global asteroseismic parameters. These were the average large
frequency separation $\left< \Delta\nu_{nl} \right>$ and the frequency
of maximum oscillation power, $\nu_{\rm max}$. Here, we have performed
a more detailed analysis of the asteroseismic data, using the
frequencies of 30 individual p modes spanning 11 radial
overtones. Using individual frequencies in the analysis enhances our ability to infer stellar properties from the seismic data.  We place much more robust (and tighter) constraints on the age than is possible from using the global parameters alone.

We estimated the frequencies of the observable p modes with a ``peak-bagging'' MCMC analysis, as per the procedures discussed in detail by \cite{Chaplin13} [see also \citealt{Carter12}]. This Bayesian machinery, in addition to allowing the use of relevant  priors, allows us to estimate the marginalized probability density function for each of the power spectrum model parameters. These parameters include frequencies, mode heights, mode lifetimes, rotational splitting, and inclination. This method performs well in low signal-to-noise conditions and returns reliable confidence intervals provided for the fitted parameters. We obtained a  best-fitting model to the observed oscillation spectrum through
optimization with an MCMC exploration of parameter space with Metropolis Hastings sampling (e.g., see \citealt{Appourchaux11}, \citealt{Handberg11}). 

We extracted estimates of the frequencies from their marginalized posterior parameter distributions. The best-fit values themselves correspond to the medians of the distributions. Table~\ref{tab:freqs} lists the estimated frequencies, along with the positive and negative 68\,\% confidence intervals. We chose not to list the estimated frequency for
the $l=2$ mode at $\simeq 3274\,\rm \mu Hz$. This mode is possibly compromised by the presence of a prominent noise spike in the power spectrum, and the posterior distribution for the estimated frequency is bimodal. The spike may be a left-over artifact of the transit, given that the frequency of the spike lies very close to a harmonic of
the planetary period. We note that omitting the estimated frequency of this mode in the modeling does not alter the final solution for the star. The lowest radial orders that we detect for the l=0 and l=1 modes are not sufficiently prominent for $l=2$ to yield a good constraint on the frequency. We exclude them from the table.

\begin{figure*}
\epsscale{0.8}
\plotone{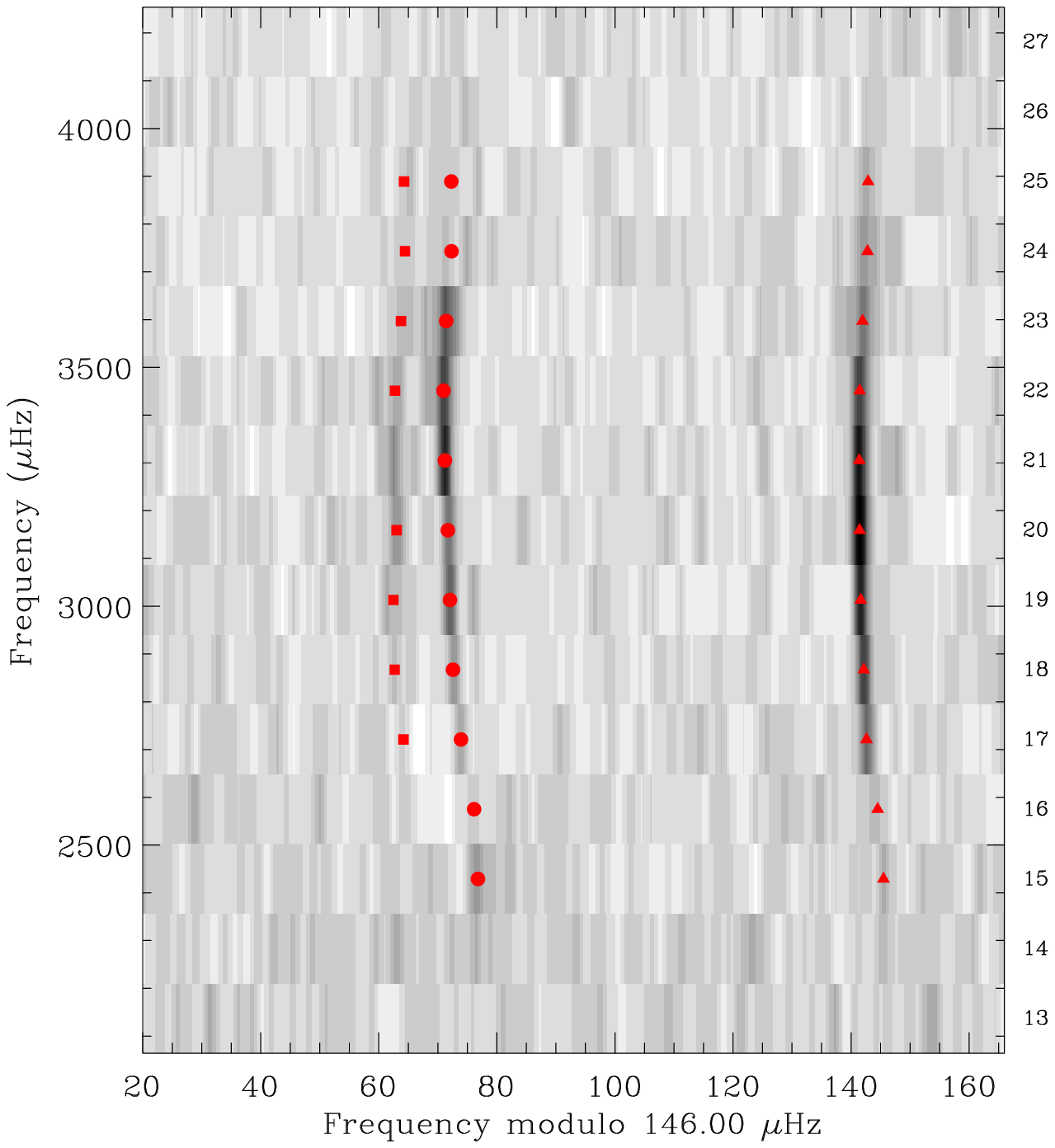}

\caption{\'Echelle diagram of the oscillation spectrum of Kepler-93. The
  spectrum was smoothed with a 1.5-$\rm \mu Hz$ Gaussian filter. The
  red symbols mark the best-fitting frequencies given by the
  peak-bagging analysis: $l=0$ modes are shown as filled circles,
  $l=1$ modes as filled triangles, and $l=2$ modes as filled
  squares. The scale on the right-hand axis marks the overtone numbers
  $n$ of the radial modes.}

\label{fig:echelle}
\end{figure*}

Fig.~\ref{fig:echelle} is an \'echelle diagram of the oscillation spectrum.  We produced this figure by dividing the power spectrum into equal segments of length equal to the average large frequency separation. We arranged the segments vertically, in order of ascending frequency. The diagram shows clear ridges, comprising overtones of each degree, $l$. The red symbols mark the locations of the best-fitting frequencies returned by the analysis described above (with $l=0$ modes shown as filled circles, $l=1$ as filled triangles and $l=2$ as filled squares).

%%%%%%%%%%%%%%%%%%%%%%%%%%%%%%%%%%%%%%%%%%%%%%%%%%%%%%%%%%%%%%%%%%%%%%%

%%%%%%%%%%%%%%%%%%%%%%%%%%%%%%%%%%%%%%%%%%%%%%%%%%%%%%%%%%%%%%%%%%%%%%%

%%%%%%%%%%%%%%%%%%%%%%%%%%%%%%%%%%%%%%%%%%%%%%%%%%%%%%%%%%%%%%%%%%%%%%%%%%%%%%%%%%%%%%%

\begin{deluxetable}{ccc}
\tabletypesize{\scriptsize} \tablecaption{Estimated oscillation frequencies of Kepler-93 (in $\rm \mu Hz$)} 
\tablewidth{0pt}
\tablehead{
 \colhead{$l=0$}& \colhead{$l=1$}& \colhead{$l=2$}}
 \startdata
$2412.80 \pm 0.36/0.29$& $2481.50 \pm 0.33/0.33$&          ...           \\
$2558.14 \pm 0.84/0.96$& $2626.52 \pm 1.57/1.15$& $2692.18 \pm 0.80/0.64$\\
$2701.92 \pm 0.18/0.18$& $2770.62 \pm 0.23/0.25$& $2836.70 \pm 0.52/0.48$\\
$2846.56 \pm 0.11/0.11$& $2916.15 \pm 0.11/0.10$& $2982.49 \pm 0.25/0.27$\\
$2992.05 \pm 0.10/0.10$& $3061.65 \pm 0.09/0.10$& $3129.02 \pm 0.27/0.31$\\
$3137.70 \pm 0.14/0.12$& $3207.45 \pm 0.10/0.09$&          ...           \\
$3283.19 \pm 0.10/0.08$& $3353.39 \pm 0.10/0.10$& $3420.73 \pm 0.34/0.51$\\
$3428.96 \pm 0.13/0.14$& $3499.46 \pm 0.16/0.14$& $3567.75 \pm 0.65/0.70$\\
$3575.41 \pm 0.24/0.24$& $3645.95 \pm 0.25/0.28$& $3714.43 \pm 1.28/1.20$\\
$3722.32 \pm 1.24/1.08$& $3792.80 \pm 0.59/0.68$& $3860.28 \pm 1.69/1.86$\\
$3868.28 \pm 0.97/1.20$& $3938.88 \pm 0.89/0.73$&          ...           \\
\enddata
\label{tab:freqs}
\end{deluxetable}

%%%%%%%%%%%%%%%%%%%%%%%%%%%%%%%%%%%%%%%%%%%%%%%%%%%%%%%%%%%%%%%%%%%%%%%%%%%%%%%%%%%%%%%
%\caption{We list error bars enclosing both upper and lower limits to 68\%}
%confidence intervals. In some cases, these error bars are}
%asymmetric. The  $l=2$ overtone at 3274 $\mu$Hz is not}
%listed because of a prominent noise spike in the power spectrum, as}
%described in the text.}

We require a complementary spectral characterization, in tandem with asteroseismology, to determine the stellar properties of Kepler-93. From a spectrum of the star, we measure its effective temperature, $T_{\rm eff}$, and its metallicity, [Fe/H]. We use these initial values, together with the two global asteroseismic parameters, to estimate the surface gravity, log\,$g$. Then, we repeat the spectroscopic analysis with log\,$g$ fixed at the asteroseismic value, to yield the revised values of $T_{\rm eff}$ and [Fe/H].  We use the spectroscopic estimates of \cite{Huber13}. These values came from analysis of high-resolution optical spectra collected as part of the \emph{Kepler} Follow-up Observing Program (KFOP). To recap briefly: \cite{Huber13}, using the Stellar Parameter Classification (SPC) pipeline \citep{Buchave12},  analyzed spectra of Kepler-93 from three sources. These were the HIRES spectograph on the 10-m Keck telescope on Mauna Kea, the FIber-fed Echelle Spectrograph (FIES) on the 2.5-m Nordic Optical Telescope (NOT) on La Palma, and the Tull Coud\'e Spectrograph on the 2.7-m Harlan J. Smith Telescope at the McDonald Observatory. In that work, they employed an iterative procedure to refine the estimates of the spectroscopic parameters (see also \citealt{Bruntt12}, \citealt{Torres12}). They achieved convergence of the inferred properties (to within the estimated uncertainties) after just a single iteration.

Five members of the team (S.~Basu, J.C.-D., T.S.M., V.S.A. and D.S.) independently used the individual oscillation frequencies and spectroscopic parameters to perform a detailed modeling of the host star. We adopted a methodology similar to that applied previously to the exoplanet host stars Kepler-50 and Kepler-65 (see \citealt{Chaplin13}; and references therein). Full details may be found in the Appendix. The final properties presented in Table 2 are the solution set provided by J.C.-D. (whose values lay close to the median solutions). We include a contribution to the uncertainties from the scatter between all five sets of results. This was done by taking the chosen modeler's uncertainty for each property and adding (in quadrature) the standard deviation of the property from the different modeling results. Our updated properties are in good agreement with---and, as expected, more tightly constrained than---the properties in \cite{Huber13}.

We note the precision in the estimated stellar properties, in particular the age, which we estimate to slightly better than 15\,\%. We estimate the central hydrogen abundance of Kepler-93 to be approximately $31\,\%$, which is similar to the estimated current value for the Sun. The age of Kepler-93 at central hydrogen exhaustion is predicted to be around 12.4\,Gyr.

%%%%%%%%%%%%%%%%%%%%%%%%%%%%%%%%%%%%%%%%%%%%%%%%%%%%%%%%%%%%%%%%%%%%%%%%%%%%%%%%%%%%%%%

%\begin{deluxetable}{ccccccc}
%\tabletypesize{\scriptsize} \tablecaption{Properties of Kepler-93} 
%\tablewidth{0pt}
%\tablehead{
% \colhead{$T_{\rm eff}$}& \colhead{[Fe/H]}& 
% \colhead{$M$}& \colhead{$R$}& \colhead{$\left< \rho \right>$}&
% \colhead{$\log\,g$}& \colhead{Age}\\ 
% \colhead{(K)}& \colhead{(dex)}&
% \colhead{($\rm M_{\odot}$)}& \colhead{($\rm R_{\odot}$)}& \colhead{($\rm g\,cm^{-3}$)}&
% \colhead{(dex)}& \colhead{(Gyr)}}
% \startdata
% $5669 \pm 75$& $-0.18 \pm 0.10$&
% $0.911 \pm 0.033$& $0.919 \pm 0.011$& $1.652 \pm 0.006$& $4.470 \pm 0.004$& 
% $6.6 \pm 0.9$\\
%\enddata
%\label{tab:res}
%\end{deluxetable}

%%%%%%%%%%%%%%%%%%%%%%%%%%%%%%%%%%%%%%%%%%%%%%%%%%%%%%%%%%%%%%%%%%%%%%%%%%%%%%%%%%%%%%%

\subsection{Derivation of Planetary Parameters}
We estimated the uncertainty in the planetary transit parameters using
the Markov Chain Monte Carlo (MCMC) method as follows. To fit the
transit light curve shape, we removed the effects of baseline drift by individually normalizing each transit. We fitted a linear function of time to the flux immediately before and after transit (specifically, from 8.6 hours to 20 minutes before first contact, equal to 3 transit durations, and an equivalent time after fourth contact).  We employed model
light curves generated with the routines in \cite{Mandel02}, which
depend upon the period $P$, the epoch $T_{c}$, the planet-to-star
radius ratio $R_{p}/R_{\star}$, the ratio of the semi-major axis to
the stellar radius $a/R_{\star}$, the impact parameter $b$, the
eccentricity $e$, and the longitude of periastron, $\omega$. We allowed
two quadratic limb-darkening coefficients (LDCs), $u_{1}$ and $u_{2}$,
to vary. Our data set comprises 233 independent transits of
Kepler-93b, all observed in short cadence mode (with exposure time 58.5 s, per \citealt{Gilliland10}). We employed the \cite{Borucki11} transit parameters 
 from the first four months of observations as a starting point to refine the ephemeris. We allowed the time of transit to float for each
individual transit signal, while fixing all other transit parameters
to their published values. We then fitted a linear ephemeris to this set
of transit times, and found an epoch of transit $T_{0}$= BJD 2454944.29227 $\pm$0.00013 and a 
period of 4.72673978$\pm$9.7$\times$10$^{-7}$ days. We iterated this
process, using the new period to fit the transit times, and found that
it converged after one iteration.

We imposed an eccentricity of zero for the light curve fit,
which we justify as follows. The circularization
time (for modest initial $e$) was reported by \cite{Goldreich66}, where $a$
is the semimajor axis of the planet, $R_{p}$ is the planetary radius,
$M_{p}$ is the planetary mass, $M_{\star}$ is the stellar mass, $Q$ is
the tidal quality factor for the planet and $G$ is the gravitational constant:

\begin{equation}
t_{\mbox{circ}}=\frac{4}{63}\frac{1}{\sqrt{GM_{\star}^{3}}}\frac{M_{p}a^{13/2}Q}{R_{p}^{5}}.
\end{equation}

\noindent Planetary $Q$ is highly uncertain, but we estimate a $Q$=100 with the assumption of a terrestrial composition. This value is typical for rocky planets in our Solar Systems \citep{Yoder95, Henning09}. We find that the circularization timescale would be 70 Myr for a 3.8 $M_{\oplus}$ planet orbiting a
0.91 $M_{\odot}$ star at 0.053 AU (we justify this mass value in Section \ref{sec:rv}). To achieve a circularization timescale comparable to the age of the star, $Q$ would
have to be on the order of $10^{4}$, similar to the value of Neptune. As we discuss in Section \ref{sec:composition}, there is only a
3\% probability that Kepler-93b has maintained an extended atmosphere. This finding renders a Neptune-like $Q$ value correspondingly unlikely. For this reason, we assume that sufficient circularization timescales have elapsed to make the $e=0$ assumption valid.

To fit the orbital parameters, we employed the Metropolis-Hastings MCMC
algorithm with Gibbs sampling (described in detail for astronomical
use in \citealt{Tegmark04} and in \citealt{Ford05}). We
first conducted this analysis using only the \kepler\
light curve to inform our fit. We incorporated the
presence of correlated noise with the \cite{Carter09} wavelet
parametrization. We fitted as free parameters the white and red 
contributions to the noise budget, $\sigma_{w}$ and $\sigma_{r}$. The allowable range of stellar density we inferred from the light curve alone is much broader
than the range of stellar densities consistent with our asteroseismic study. Crucially, they are consistent. When we fit the transit parameters independently of any knowledge of the
star (allowing $a/R_{\star}$ to float), we found that values of $a/R_{\star}$ from
11.93--12.78 gave comparable fits to the light curve. This parameter varies
codependently with the impact parameter (the ingress and
egress times can vary within a 30 s range, from 2.5--3 minutes). We then refitted
the transit light curve and apply the independent asteroseismic
density constraint as follows. Rearranging Kepler's Third Law in the manner
employed by \cite{Seager03}, \cite{Sozzetti07} and \cite{Torres08}, we
converted the period $P$ (derived from photometry) and the stellar
density $\rho_{\star}$, to a ratio of the semi-major axis to the
radius of the host star, $a/R_{\star}$:

\begin{equation}
\left(\frac{a}{R_{\star}}\right)^{3}=\frac{GP^{2}(\rho_{\star}+p^{3}\rho_{p})}{3\pi}
 \label{eq:kepler_5}
\end{equation}
\noindent where $p$ is the ratio of the planetary radius to the
stellar radius and $\rho_{p}$ is the density of the planet. We will
hereafter assume $p^{3}\ll1$ and consider this term negligible. We mapped the
stellar density constraint, $\rho_{\star}$=1.652$\pm$0.006 g cm$^{-3}$, to
$a/R_{\star}$=12.496$\pm$0.015, which we apply as a Gaussian prior in the light curve fit.

\begin{figure}%[h!]
\begin{center}
 \includegraphics[width=5in]{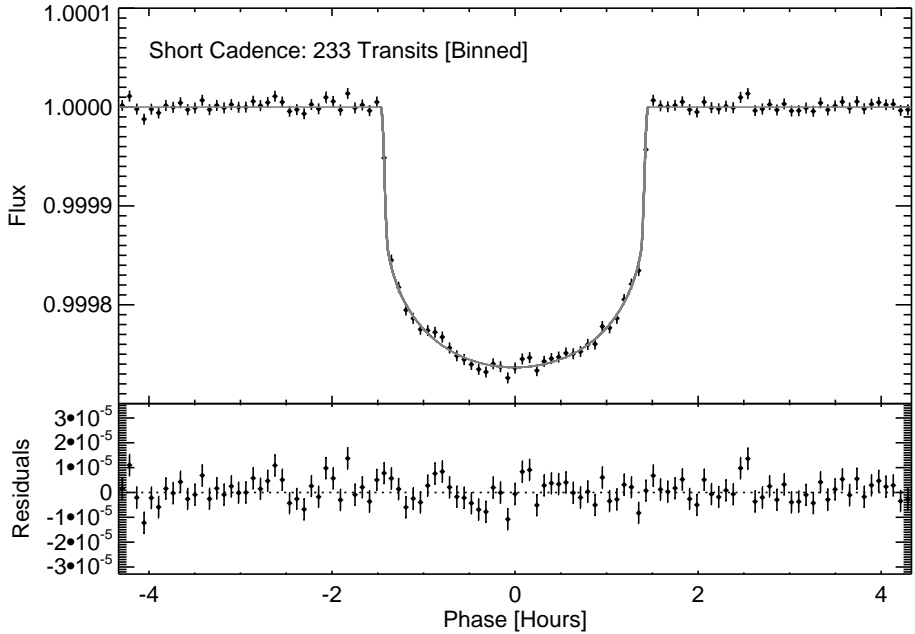} 
 \caption{Kepler-93b transit light curve as a function of planetary
   orbital phase for Quarters 1--12, gathered in short-cadence observing mode. The best-fit transit model is overplotted in grey, with residuals to the fit shown in the bottom panel.}
  \label{fig:keplerfit}
\end{center}
\end{figure}

\begin{figure}%[h!]
\begin{center}
 \includegraphics[width=8in,angle=90]{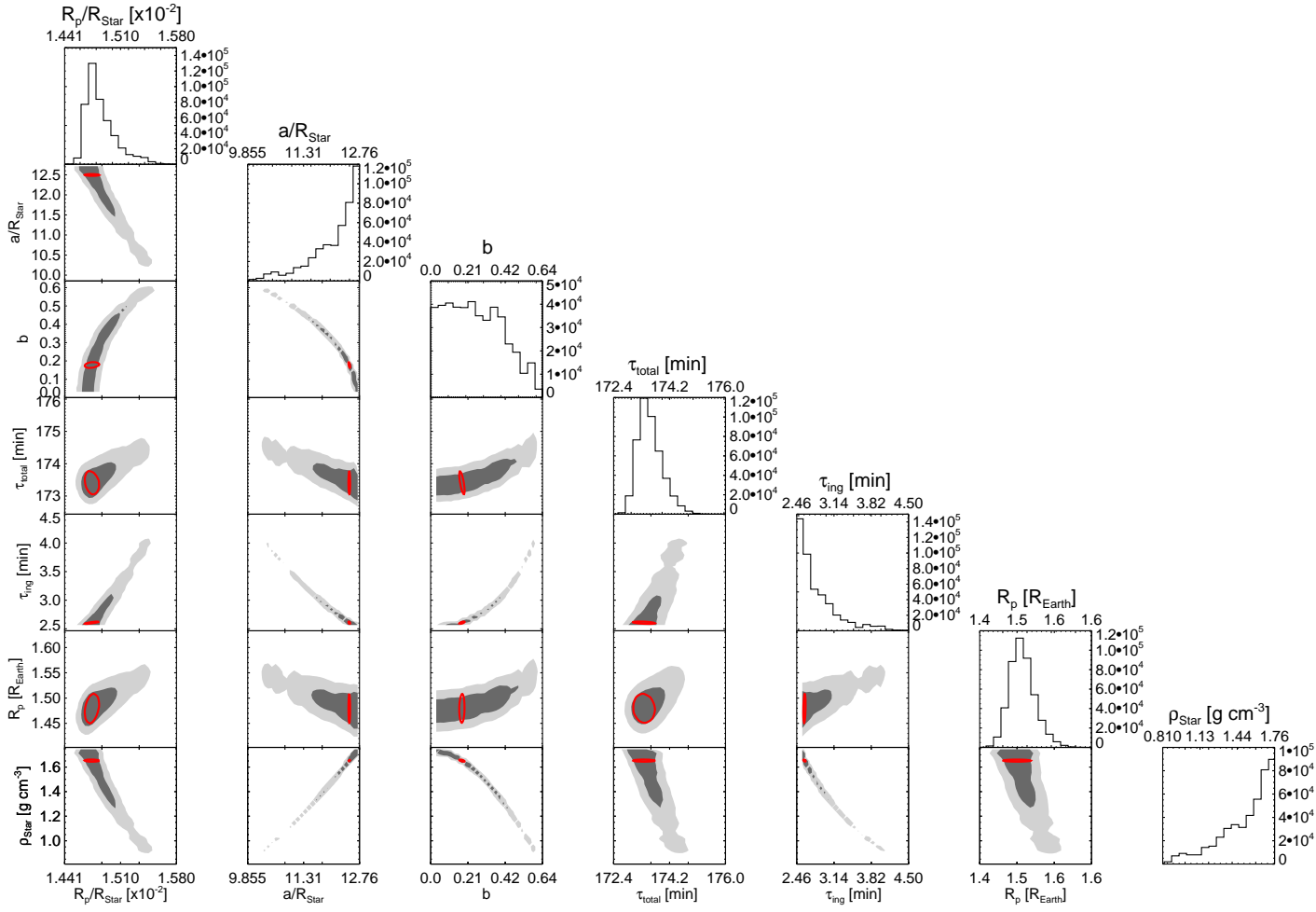} 
 \caption{Markov Chain Monte Carlo probability distributions for light
   curve parameters of Kepler-93b. The dark gray area encloses 68\% of
   the values in the chain, while the light grey area encloses 95\% of
   the values. We assign the range of values corresponding to
   1$\sigma$ confidence from the area enclosing 68\% of the values
   nearest to the mode of the posterior distribution for each
   parameter (as described in the text. When we impose the prior on
   $a/R_{\star}$ derived from asteroseismology, the 1$\sigma$ confidence contours are revised to the area depicted in red.}
   \label{fig:mcmc_results2}
\end{center}
\end{figure}

In Figure \ref{fig:keplerfit}, we show the phased \kepler\ transit light curve for Kepler-93b, with the best-fit
transit light curve overplotted. In Figure \ref{fig:mcmc_results2}, we show the correlations between
the posterior distributions of the subset of parameters in the model fit,
as well as the histograms corresponding to each parameter. We depict two cases. In one case, we allowed the stellar density to float, in the other we applied a Gaussian prior to force its consistency with the
asteroseismology analysis. We report the best-fit parameters and
uncertainties in Table \ref{tbl:res}. We determined the range of acceptable
solutions for each of the light curve parameters as follows. In the same manner as \cite{Torres08}, we report the most likely value
from the mode of the posterior distribution, marginalizing over all
other parameters. We quote the extent of the posterior distribution that encloses 68\% of values closest to the
mode as the uncertainty. To estimate an equilibrium
temperature for the planet, we assumed that the planet possesses a Bond albedo ($A_{B}$) of 0.3 and that it radiates energy equal to the energy incident upon it.  To generate physical quantities for the radius of the planet
$R_{p}$, we multiplied the $R_{p}/R_{\star}$ posterior distribution by the posterior on $R_{\star}$. In the case where we applied the asteroseismic prior, we found a planetary radius of 1.481$\pm0.019$ $R_{\oplus}$ with 1$\sigma$ confidence. This value includes the uncertainty in stellar radius. We note that the uncertainty of $R_{p}/R_{\star}$ makes up 30\% of the error budget on the planetary radius. Uncertainty on $R_{\star}$ itself contributes the remaining 70\%. Without the asteroseismic prior on $a/R_{\star}$ (and the correspondingly less constrained measurement of  $R_{p}/R_{\star}$), our measurement of the planetary radius is 1.485$\pm$ 0.27 $R_{\oplus}$. The contributions are nearly reversed in this case: 68\% of the error in $R_{p}$ is due to uncertainty in $R_{p}/R_{\star}$ and only 32\% to uncertainty in the stellar radius. 

The uncertainty in planetary radius of 0.019 $R_{\oplus}$ corresponds to 120 km. Kepler-93b is currently the planet with the most precisely measured extent outside of our Solar System. A reanalysis of \kepler's first rocky exoplanet Kepler-10b \citep{Batalha11} by \cite{Fogtmann14} with 29 months of \kepler\ photometry returned a similarly precise radius measurement: 1.460$\pm$0.020 $R_{\oplus}$. Kepler-16b \citep{Doyle11}, with radius 8.449$^{+0.028}_{-0.025}$ $R_{\oplus}$ and Kepler-62c
\citep{Borucki13}, with radius 0.539$\pm$0.030 $R_{\oplus}$, are the next most precisely measured exoplanets within the published literature. The sample of these four currently comprises the only planets outside the Solar System with radii measured to within 0.03 $R_{\oplus}$.

\begin{deluxetable*}{rrr}
\tabletypesize{\scriptsize}
\singlespace
%\tablenum{1}
\tablecaption{Star and Planet Parameters for Kepler-93}
\label{tbl:params_5}
\tablewidth{0pt}
\tablehead{
\colhead{Parameter} & \colhead{Value \& 1$\sigma$ confidence interval} & \colhead{  } \\}
\startdata
Kepler-93 [star]& & \\
\hline
  &  &  \\
Right ascension\tablenotemark{a} & 19h25m40.39s & \\
Declination\tablenotemark{a} & +38d40m20.45s &  \\ 
$T_{\mbox{eff}}$ [K] &  5669$\pm$75 & \\
$R_{\star}$ [Solar radii] & $0.919 \pm 0.011$ & \\
$M_{\star}$ [Solar masses] &  $0.911 \pm 0.033$ & \\
$\mbox{[Fe/H]}$ & -0.18$\pm$0.10 &  \\
log(g) & $4.470 \pm 0.004$ & \\
Age [Gyr] & $6.6 \pm 0.9$ &  \\
  &  &  \\
Light curve parameters  & No asteroseismic prior & With asteroseismic prior \\
\hline
&  &  \\
$\rho$ [g cm$^{-3}$] & 1.72$^{+0.04}_{-0.28}$ & 1.652$\pm0.0060$ \\ 
Period [days]\tablenotemark{b} & 4.72673978$\pm$9.7$\times$10$^{-7}$ & -- \\
Transit epoch [BJD]\tablenotemark{b}  & 2454944.29227 $\pm$0.00013 & -- \\
$R_{p}/R_{\star}$ & 0.01474$\pm0.00017$ & 0.014751$\pm0.000059$  \\ 
a/$R_{\star}$ &        12.69$^{+0.09}_{-0.76}$ & 12.496 $\pm0.015$ \\
inc [deg] &  89.49$^{+0.51}_{-1.1}$ & 89.183$\pm0.044$ \\
$u_{1}$ &  0.442$\pm0.068$ & 0.449$\pm0.063$ \\
$u_{2}$ &   0.187$\pm0.091$ & 0.188$\pm0.089$ \\
Impact Parameter &  0.25$\pm0.17$  & 0.1765$\pm0.0095$ \\
Total Duration [min] &  173.42$\pm0.36$ & 173.39$\pm{0.23}$ \\
Ingress Duration [min] & 2.52$^{+0.37}_{-0.06}$ & 2.61$\pm0.013$ \\ 
  &  &  \\
Kepler-93b [planet] & No asteroseismic prior & With asteroseismic prior \\
\hline
&  &  \\
$R_{p}$ [Earth radii] & 1.483$\pm0.025$ &1.478$\pm0.019$\\ 
Planetary $T_{eq}$ [K] &  1039$\pm$26  &  1037$\pm$13 \\ 
$M_{p}$ [Earth masses]\tablenotemark{c} & 3.8$\pm$1.5 & -- \\ 
\enddata
\label{tbl:res}
\tablenotetext{a}{ICRS (J2000) coordinates from the TYCHO reference catalog \citep{Hog98}. The proper motion derived by \cite{Hog00} -26.7$\pm$1.9 in right ascension and -4.4$\pm$1.8 in declination.}
\tablenotetext{b}{We fit the ephemeris and period of Kepler-93 in an iterative fashion, as described in the text. We fix the other orbital parameters to fit individual transit times, and report the best linear fit to these times, rather than simultaneously fitting the transit times while imposing the asteroseismic prior.}
\tablenotetext{c}{\cite{Marcy14} describe the Kepler-93 radial velocity campaign. The stated value here is a revision, with an additional year of observations, of their published value of 2.6$\pm$2.0 $M_{\oplus}$.}
\end{deluxetable*}

\section{{\it Spitzer} Observations and Analysis}
We gathered 4 transits of Kepler-93b with Warm \spitzer\ as part of program 60028 (PI: Charbonneau). These observations spanned 7.5 hours each. They are centered on the time of predicted transit, with 2 hours of out-of-transit baseline observations before ingress and after egress. The \spitzer\ Science Center, as part of IRAC calibration program 1333, observed Kepler-93 an additional 6 times to test the efficacy of the "peak-up" observational technique. This technique employs the Spitzer Pointing Control Reference Sensor (PCRS), on either the science target or a nearby calibrator target, to guide the science target within 0.1 pixels of the peak sensitivity location. These observations spanned 6 hours each. They are similarly centered on transit, with out-of-transit baselines of 1.5 hours preceding and following transit. We gathered all frames at 4.5 $\mu$m. In this channel, the intrapixel sensitivity effect is lessened as compared to 3.6 $\mu$m \citep{Ingalls12}. These observations comprise 10 total visits, of which we exclude 3 with the following reasoning. We do not include here the first two observations of Kepler-93 as part of program 60028 (Astronomical Observation Requests, or AORs, 39438336 and 39438592), which we gathered with exposure time of 10 s in the full array observing mode. These frames heavily saturated due to the 10s exposure time, and so are not useful test observations for the precision of IRAC. For all subsequent observations, we employed the 2 s exposure time and the sub-array mode, centered upon a 32$\times$32 section of the IRAC detector. Neither do we include in our analysis the first pointing of the \spitzer\ Science Center to Kepler-93 (AOR 44448000). For these observations, the spacecraft pointing control relied upon gyroscopes rather than the usual star trackers. The pixel drift for these observations was unusually large, spanning several pixels, as compared to the usual 0.1 pixels. The coarse sampling of these observations does not allow meaningful removal of the pixel systematic effects. The remaining 7 transits span 2010 November through 2011 November.  We have listed all of the AORs associated with these transit observations of Kepler-93 in Table \ref{tbl:spitzer}. 

The limiting systematic error for time-series photometry with post-cryogenic IRAC is the variable spatial sensitivity of the pixel over the typical range of pointing drift. The undersampling of the point-spread function results in the ``pixel-phase'' effect (see eg. \citealt{Charbonneau05, Knutson08}). The star appears brighter or dimmer as the core of the PSF lies on more or less sensitive portions of the same pixel. Typical \spitzer\ photometric observations can display variations in brightness of 8\% for a star with instrinsically constant brightness \citep{Ingalls12}. The optimal means of fitting and removing the intrapixel sensitivity variation to uncover the astrophysical signal is the subject of many papers
(\citealt{Charbonneau05, Knutson08, Ballard10b, Demory11b, Stevenson12}, among others). The traditional method for removing the pixel-phase effect is to model the variation as a polynomial in $x$ and $y$. Other approaches include those of \cite{Demory11b}, in which the authors included terms in their pixel model that depend upon time. Both the methods of \cite{Ballard10b} and \cite{Stevenson12} do not rely on a particular functional form for the intrapixel sensitivity, but rather model the sensitivity behavior with a weighted sum of the brightness measurements. These weighted sums are either performed individually for each measurement (as in the \citealt{Ballard10b} methodology), or interpolated onto a grid with resolution that maximizes the final precision of the light curve (as in the BLISS mapping of \citealt{Stevenson12}). 

% When \spitzer\ pointing drift occurs over a larger region of the pixel, the map of its spatial variation is coarser. 

We describe our particular technique for processing and extracting photometry from IRAC images in \cite{Ballard11b} and \cite{Ballard13}. We reduced all of these data in a uniform fashion, as described in these works. We deviate here in only one way from this published process, in order to account differently for correlated noise. We include in our transit fit model the red and white noise coefficients associated with the wavelet correlated noise model developed by \cite{Carter09}.  The Transit Analysis Program (TAP, \citealt{Gazak12}) incorporates this machinery, and we used it to fit our transits. This model includes the parameters $\sigma_{r}$ and $\sigma_{w}$, which reflect the contributions of red (correlated) and white (Gaussian) noise budgets to the photometry. The short out-of-transit baseline associated with typical \spitzer\ transit observations makes it difficult to use the alternative residual permutation treatment \citep{Winn09}. That method relies upon having enough observations to sample correlated noise on the timescale of the transit.

\begin{figure}[h!]
\begin{center}
 \includegraphics[width=7in]{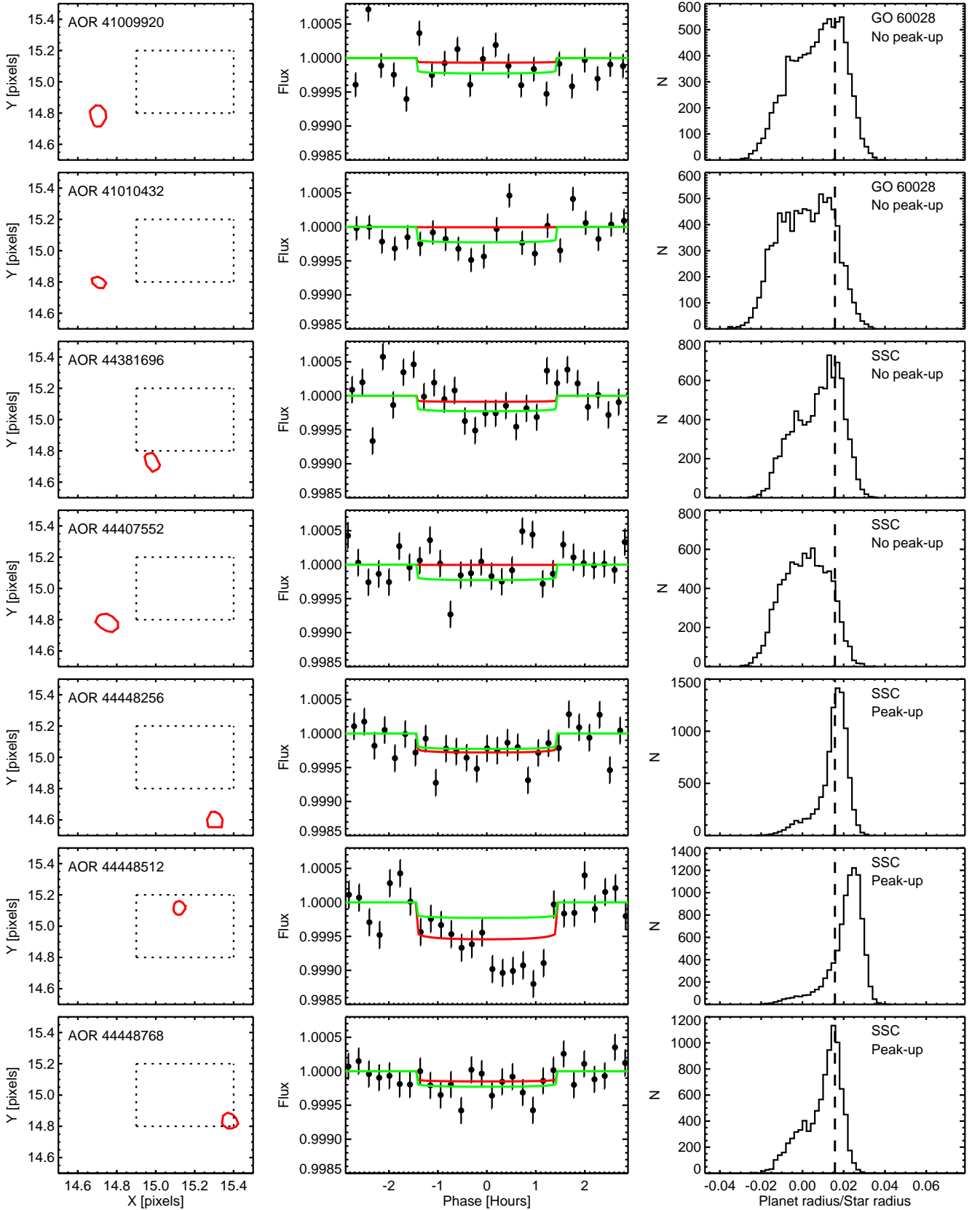} 
 \caption{{\it Left panels:} Location on the IRAC detector of the
   observations of Kepler-93. We gathered observations at the central pixel
   of the 32$\times$32 subarray of IRAC. The red contour encloses 68\%
   of centroids. We have indicated with a dotted line the approximate area which was mapped in detail with the standard star BD+67 1044 with the intent of producing a pixel flat-field \citep{Ingalls12}. {\it Middle panels:} Transits of Kepler-93b,  binned
   to 15 minute cadence. The best-fit transit model with depth derived
   from the \spitzer\ observations is shown with a solid red line,
   while the \kepler\ transit model (with \spitzer\ 4.5 $\mu$m channel
   limb darkening) is shown in green. {\it Right panels:} The
   posterior distribution on $R_{p}/R_{\star}$ on each \spitzer\ transit. The depth measured in the \kepler\ bandpass is marked with a dashed line. The mode of pointing is noted in the upper left hand corner, as well as the program associated with the AOR.}
  \label{fig:spitzer_5}
\end{center}
\end{figure}

In Figure \ref{fig:spitzer_5}, we show the transit light curves and $R_{p}/R_{\star}$ posterior distributions for each transit. We also mark the spatial locations of the observations on IRAC detector. We indicate the approximate region well-mapped with a standard star by the Spitzer Science Center \citep{Ingalls12}. Only two of the four transit observations gathered with peak-up mode fell mostly within this well-mapped region, for which there exist publicly available pixel maps \citep{Ingalls12}. For this reason, we defaulted to the traditional self-calibration technique. For AOR 4448512, we did employ the pixel map for 4.5 $\mu$m and found similar photometric precision and the same transit depth. Therefore, we employed self-calibration uniformly for all transits. We report the measured \spitzer\ values for $R_{p}/R_{\star}$ of Kepler-93b in Table \ref{tbl:spitzer}. We list the red and white noise coefficients, $\sigma_{r}$ and $\sigma_{w}$, at the 10s timescale for each transit. We also report the ratio between these contributing noise sources on both the 10s and 3 hour (transit) timescales.

We note that using peak-up mode in the {\it Spitzer} observations of Kepler-93b improves the precision with which we measure the transit radius ratio by a factor of two. The average error bar on $R_{p}/R_{\star}$ before peak-up is 1.5$\times$10$^{-2}$. After peak-up, the average error is is 7.3$\times$10$^{-3}$. Since the transit radius ratio itself is only 1.472$\times$10$^{-2}$, we have detected the transit event in individual \spitzer\ light curves with 2$\sigma$ certainty when peak-up is implemented. The improvement is reflected in the reduced values of $\sigma_{r}$, which encodes the presence of correlated noise in the observations: it is lower by a factor of 1.5--3. We understand the narrower posterior distributions of $R_{p}/R_{\star}$ associated with peak-up mode to be less inflated by the presence of red noise. The ratio of red/white noise for the transit duration is 0.3 with peak-up, and 0.8 without. While correlated noise is still present in the peak-up observations, it is significantly reduced.

If we combine all seven measurements of Kepler-93b gathered by \spitzer\ at 4.5 $\mu$m, then we find a planet-to-star radius ratio of $R_{p}/R_{\star}$=0.0144$\pm$0.0032. If we consider only the values gathered with peak-up mode enabled (that is, if we use only the latter three of the seven transits) we find $R_{p}/R_{\star}$ of 0.0177$\pm0.0037$, which is comparable precision with half the observing time.

\begin{deluxetable*}{lclccccc}
\tabletypesize{\scriptsize}
\singlespace
%\tablenum{1}
\tablecaption{Warm \spitzer\ Observations of Kepler-93b}
\label{tbl:spitzer}
\tablewidth{0pt}
\tablehead{
\colhead{AOR} & \colhead{Date of Observation} &\colhead{Observing
  mode}  & \colhead{$R_{p}/R_{\star}$}   &
\colhead{$\sigma_{w}$\tablenotemark{a} [$\times10^{-3}$]} &
\colhead{$\sigma_{r}$\tablenotemark{a} [$\times10^{-3}$]} &
$\alpha_{\mbox{[10 s]}}$ & $\alpha_{\mbox{[transit]}}$\\}
\startdata
41009920 & 2010 November 11 & No peak-up\tablenotemark{b} &
 0.008 $^{+0.012}_{-0.015}$ & 1.46 & 0.24 & 0.14 & 0.83 \\
41010432 & 2010 December 09 & No peak-up\tablenotemark{b} &
 0.002 $^{+0.013}_{-0.014}$ &1.42 & 0.24 & 0.15 & 0.83 \\
44381696 & 2011 September 19 & No peak-up\tablenotemark{c} &
 0.0092 $^{+0.0095}_{-0.014}$ & 1.55 & 0.20 & 0.11 & 0.80 \\
44407552 & 2011 October 07 & No peak-up\tablenotemark{c} 
& 0.002 $^{+0.011}_{-0.012}$ & 1.64& 0.20 & 0.10 & 0.78\\
44448256 & 2011 October 26 & Peak-up\tablenotemark{c}  & 0.0163
$^{+0.0046}_{-0.0080}$ & 1.67 & 0.026 & 0.016 & 0.32\\
44448512 & 2011 October 31 & Peak-up\tablenotemark{c}  & 0.0228
$^{+0.0053}_{-0.0090}$ & 1.62 & 0.15 & 0.083 & 0.73\\
44448768 &  2011 November 05 & Peak-up\tablenotemark{c} & 0.0119 $^{+0.0060}_{-0.013}$ & 1.64 & 0.018 & 0.011 & 0.26\\
\enddata
\label{tbl:spitzer}
\tablenotetext{a}{As defined in \cite{Carter09}, for 10 s bin size.}
\tablenotetext{b}{Data gathered as part of GO program 60028 (P.I. Charbonneau).}
\tablenotetext{c}{Data gathered as part of the IRAC Calibration program 1333 by the \spitzer\ Science Center.}
\end{deluxetable*}

\section{False-Positive Tests for Kepler-93b}

Both large-scale studies of the \kepler\ candidate sample, and individual studies of Kepler-93 specifically, consider its false positive probability. \cite{Morton11} provided {\it a priori} false positive probabilities for the \kepler\ planetary candidates published by \cite{Borucki11}. They include Kepler-93 in their sample. They cite the vetting of candidates by the \kepler\ software (detailed by \citealt{Batalha10}) as being already sufficient to produce a robust list of candidates. They combined stellar population synthesis and galactic structure models to conclude a generally low false-positive rate for KOIs. Nearly all of the 1235 candidates they considered in that work have a false positive probability $<$10\%.  Kepler-93, with a \kepler\ magnitude of 9.93 and a galactic latitude of 10.46$^{\circ}$, has an  {\it a priori} false positive probability of 1\%. \cite{Fressin13} considered false positive rates as a function of planetary size. The false positive rate for the radius bin relevant to Kepler-93b (1.25--2 $R_{\oplus}$) is 8.8$\pm$1.9\%. \cite{Marcy14} consider Kepler-93b specifically. They leverage the false positive machinery described by \cite{Morton12} (which employs the transit light curve, spectroscopy, and adaptive-optics imaging) to infer a false positive probability of $<10^{-4}$. We consider the false positive probability below from the angle of stellar density, which is probed independently by asteroseismology and by the shape of the transit. We also rule out a remaining false positive possibility broached by \cite{Marcy14}, based upon the radial velocity trend they observed.

\subsection{Asteroseismic and Photometric Stellar Density Constraints}
\label{sec:rho_fpp}

The transit light curve alone provides some constraints on the host star of the planet. False-positive scenarios in which an eclipsing binary comprising (a) two stars, or (b) a star and a planet, falls in the same aperture as the \kepler\ target star, will produce a transit signal with a diluted depth. A comparison of the detailed shape of the transit light curve to models of putative blend scenarios, as described by \cite{Torres04},  \cite{Fressin11}, and \cite{Fressin12}, constrains the parameter space in which such a blend can reside. The
likelihood of such a scenario, given the additional observational constraints of spectra and adaptive optics imaging, is then weighed against the likelihood of an authentic planetary scenario. This practice returns a robust false-positive probability. Here, we rather comment on the consistency of the transit light curve with our hypothesis that it originates from a 1.48 $R_{\oplus}$ planet in orbit around a 0.92 $R_{\odot}$ star. 

The {\it period} and {\it duration} of the transit will not be affected by the added light of another star. The steepness of the ingress and egress in the transit signal of Kepler-93b (enabled by the wealth of short-cadence data on the star), which lasts between 3.0 and 2.46 minutes as we report in Table \ref{tbl:res}, reasonably precludes a non-planetary object. The transiting object passes entirely onto the disk of the star too quickly to be mimicked by an object that subtends a main sequence stellar radius. We note that a white dwarf could furnish such an ingress time, but would produce a radial velocity variation at the 4.7 day period at the level of 100 km/s,  much larger than we observe (we describe the radial velocity observations in the following subsection). We conclude that a blend scenario (a) comprising three stars is unlikely. False positive scenario (b) involves a transiting planet system within the \kepler\ aperture that is not in orbit around the brightest star. Rather, the diluted transit depth would conspire to make the planet appear smaller than it is in reality. In this case, transit duration still constrains the density of the host star. Certain stellar hosts are immediately ruled out, because a planet orbiting them could not produce a transit as long as observed, given the eccentricity constraint implied by the short planetary period. We constrain the host from the transit light curve parameters as follows.

\cite{Winn09} described the approximation of the transit depth for cases in which the eccentricity $e$ is close to zero (which we assume is valid because of the short period), where
$R_{p}\ll R_{\star} \ll a$ (where $a$ is the semi-major axis of the planet's orbit) and where the impact parameter $b\ll1-R_{p}/R_{\star}$. In these cases, the transit timescale $T$,
defined to be the total duration of the transit minus an ingress time, is related to a characteristic timescale $T_{0}$ as follows:

\begin{equation}
T_{0}\approx \frac{T}{\sqrt{1-b^{2}}},
\end{equation}

where

\begin{equation}
T_{0}=\frac{R_{\star}P}{\pi a}\approx 13 \mbox{ hr }\left(
  \frac{P}{\mbox{1 yr}}  \right)^{1/3} \left(\frac{\rho_{\star}}{\rho_{\odot}} \right)^{-1/3}.
\end{equation}

\noindent  We computed the distribution for $T_{0}$ from our posterior
distributions for the total duration, ingress duration, and
impact parameter. We found that values between 170 and 184 minutes are
acceptable at the 1$\sigma$ confidence limit. This translates to a
constraint on the exoplanet's host star density of 1.72$^{+0.04}_{-0.28}$ g
cm$^{3}$, as given in Table \ref{tbl:res}. The density of the \kepler\ target star measured from asteroseismology, 1.652$\pm$0.006 g cm$^{-3}$, is within 1$\sigma$ of the value inferred from the transit. This consistency lends credence to the interpretation that the bright star indeed hosts the planet.  

\subsection{\spitzer\ and \kepler\ Transit Depths}
The transit depth we report with \spitzer\ of $R_{p}/R_{\star}$=0.0144$\pm$0.0032 lies within 1$\sigma$ of
the depth measured by \kepler\ of
0.014751$\pm$0.000057 (as we report in Table \ref{tbl:res}). The achromatic nature of the transit depth disfavors blended ``false-positive'' scenarios for
Kepler-93b and render it more likely to be an authentic planet. D\'esert et al. (in preparation) consider the \spitzer\ transit depth, the \spitzer\ magnitude at 4.5 $\mu m$, and adaptive optics imaging to find a false-positive probability of 0.18\%. 

\subsection{Spectra and Adaptive Optics Imaging}
\label{sec:rv}

\cite{Marcy14} measured the radial velocities of Kepler-93b. They employed Keck-HIRES spectra spanning 1132 days, from 2009
July to 2012 September. In that work, they describe their reduction and radial velocity fitting process in detail. While they report a mass of 2.6$\pm$2.0 $M_{\oplus}$ in that work, the addition of another 14 spectra gathered in the 2013 observing season refine this value to 3.8$\pm$1.5 $M_{\oplus}$ (Geoffrey Marcy, private communication). We employ this value, a 2.5$\sigma$ detection of the mass, for the remainder of our analyses. 

They report a linear radial velocity trend in the Keck-HIRES observations of Kepler-93. This trend is present at a level of 10 m s$^{-1}$ yr$^{-1}$ over a baseline of 3 years. They conclude that the trend is caused by another object in the Kepler-93 system, and designate it Kepler-93c. They place lower limits on both its mass and period of M$>$3M$_{JUP}$ and P$>$5 yr. At this lower mass limit the object would be orbiting in a near-transiting geometry (though no additional transits appear in the \kepler\ light curve). With an additional year of observations, the linear trend is still present. Because the radial velocities have not yet turned over, we assert that its period must in fact be larger than twice the observing baseline, which is now three years. We assume a lower limit to the period of the perturber of 6 years. There exists the possibility, which \cite{Marcy14} note in their discussion of Kepler-93, that the planet is in fact orbiting the smaller body responsible for the radial velocity drift. In this case, the planetary properties must be revisited based upon the revised stellar host. However, while the RV data alone permit this scenario, it is ruled out based upon the density inferred from the transit light curve, in tandem with other observables. We explain our reasoning here.

In Section \ref{sec:rho_fpp}, we argue that  the host star to the planet must possess a density within the range of 1.72$^{+0.04}_{-0.28}$ g cm$^{-3}$ to be consistent with the transit duration. We compared this density to the Dartmouth stellar evolutionary models given by \cite{Dotter08} to place an approximate constraint on the host star mass. We evaluated the Dartmouth models over a grid of metallicities (sampled at 0.1 dex from -1 to 0.5 dex) and a grid of ages between 500 Myr and 14 Gyr (sampled at 0.1 Gyr). We do not consider enhancement in alpha elements. We found that the 1$\sigma$ range in observed density translates to a range of acceptable stellar masses from 0.75 to 1.23 $M_{\odot}$. If we assume that this putative host star is physically associated with the target star (for reasons we explore in the following paragraph), and we further require that these two stars formed within one Gyr of each other, then the mass is constrained to between 0.82 and 1.02 $M_{\odot}$. At the 3$\sigma$ level of uncertainty, the density lies between 0.86 and 1.76 g cm$^{-3}$. The range of acceptable masses within this broader confidence interval is correspondingly 0.75--1.52 $M_{\odot}$. An assumption of contemporaneous formation of the two stars further constrains this range to 0.82--1.12 $M_{\odot}$. Therefore, we preclude stars below 0.75 M$_{\odot}$ as the planetary host. To be the perturber to the asteroseismic target star, the two stars are physically associated and we assume their formations to be coeval. A 0.75 $M_{\odot}$ star, if it possessed an age within 1 Gyr of the target star, would be between 0.7 and 1.0 magnitudes fainter in $K$ than the brighter target star. The exact difference in magnitude, though within this range, depends upon its exact metallicity and age. With a period greater than 6 years and a mass $>0.75 M_{\odot}$, its semimajor axis is greater than 3.9 AU.

The lack of a set of additional lines in the high-resolution spectrum brackets the upper limit on the mass and velocity of the object perturbing the bright star. Any star brighter than 0.3\% the brightness of the primary star is ruled out, provided the radial velocity difference between the bright star and the perturbing body is greater than 10 km s$^{-1}$. A putative star 1 magnitude fainter than the target star (40\% its brightness) is therefore required to possess a velocity within 10 km s$^{-1}$ of the target star. The overlap of absorption features precludes the detection of another star if their radial velocities are sufficiently similar. For the sake of argument, we proceed under the assumption that the additional star satisfied this condition.

\cite{Marcy14} use Keck adaptive optics images of Kepler-93 to rule out any companion beyond beyond 0.1'' within 5 magnitudes. We compute a physical scale corresponding to this angular size. We compare the observed \kepler\ magnitude of 9.931 to the magnitude predicted in the \kepler\ bandpass for a star with the mass, metallicity, and age of Kepler-93 (listed in Table \ref{tbl:res}) by the Dartmouth Stellar evolutionary models. We find a predicted \kepler\ magnitude at a distance of 10 pc of 4.9$\pm$0.1 magnitudes. We therefore estimate that the star is approximately 100 pc away from Earth, so that 0.1'' is physically equivalent to 10 AU. The minimum distance between the stars of 3.9 AU (set by the lower limit on the period of the perturber) is an angular separation of 0.039". The adaptive optics imaging rules out any object within 2 magnitudes of the host star at this separation. The putative other host star to the planet, only 1 magnitude fainter, is therefore further constrained to lie in such a geometry and phase that it eluded AO detection. However unlikely, this scenario is plausible. 

Therefore, we consider a hypothetical 0.75 M$_{\odot}$ planet-host companion to the bright star, with a semimajor axis of 3.9 AU. It possessed a velocity within 10 km s$^{-1}$ of the host star in the high-resolution spectrum that \cite{Marcy14} searched for a set of additional lines.  At the time of the adaptive optics imaging gathered by \cite{Marcy14}, it resided at a phase in its orbit where its projected distance from the brighter star rendered it invisible. It is then the several-year baseline of radial velocity observations that conclusively rule out the existence of this companion. If this companion existed at 3.9 AU, the brighter star would possess a radial velocity semi-amplitude of 15.5 km s$^{-1}$ over the 6 year baseline. Given the drift in half of this period of only 45 m s$^{-1}$, its inclination would have to be within 0.18$^{\circ}$ of perfectly face-on to be consistent. If it were indeed in such a face-on orientation, it would be readily detectable with adaptive optics imaging within the stated limits. We conclude, based upon the density constraint from the transit light curve in combination with published radial velocity and spectral constraints, that the planet indeed orbits the brighter target star. Whether or not the radial velocity trend is due to an additional Jovian planet or to a small star remains unresolved.

%\subsection{Consistency with Theoretical Limb Darkening Prediction}

%From \cite{Claret13}, given a log(g) of 4.5, $T_{\mbox{eff}}$ of 5600,
%and [M/H] of zero, $u_{1}$=0.5102 and $u_{2}$=0.1712 using the
%least-squares method of fitting the theoretical limb-darkening
%coefficients and $u_{1}$=0.4987 and $u_{2}$=0.1772 using the flux
%conservation method. We select the limb darkening coefficients
%associated with a microturbulent velocity of 2 km s$^{-1}$, which is
%the only option with the other specified parameters. We take the mean
%value for either method between the values at 5600 and 5700 K (given the
%stellar temperature of 5669$\pm$75), and overplot it against the
%confidence contours we derive on the limb-darkening coefficients from
%the light curve alone in Figure \ref{fig:ldc}. These values are 0.5044
%and 0.1742 for $u_{1}$ and $u_{2}$ using least squares, respectively,
%and 0.5285 and 0.1267 for $u_{1}$ and $u_{2}$ from flux
%conservation. From the light curve fit with no asteroseismic prior, we find $u_{1}$ of
%0.442$\pm0.039$ and $u_{2}$ of 0.178$\pm0.092$. The theoretical
%limb-darkening coefficients lie within the 2$\sigma$
%contour of the pair of quadratic limb-darkening values we infer from the light curve. 

\section{Discussion}

\subsection{Composition of Kepler-93b}
\label{sec:composition}

There now exist eight exoplanets with radii in the 1.0-1.5 $R_{\oplus}$ range with dynamically measured masses. We define the radius here as the most-likely value reported by the authors, and define a mass detection as residing more than 2 standard deviations from a mass of zero. From smallest to largest, these are Kepler-102d \citep{Marcy14}, Kepler-78b \citep{Howard13,Pepe13}, Kepler-100b \citep{Marcy14}, Kepler-10b \citep{Batalha11}, Kepler-406b \citep{Marcy14}, Kepler-99b \citep{Marcy14}, Kepler-93b, and Kepler-36b \citep{Carter12}. The mass of Kepler-93b is detected by \cite{Marcy14} at 3.8$\pm$1.5 $M_{\oplus}$. This translates to a density of 6.3$\pm$2.6 g cm$^{-3}$. Among the worlds smaller than 1.5 $R_{\oplus}$, the density of Kepler-93b is statistically indistinguishable from those of Kepler-78b, Kepler-36b, and Kepler-10b. 

\begin{figure}[h!]
\begin{center}
 \includegraphics[width=5in]{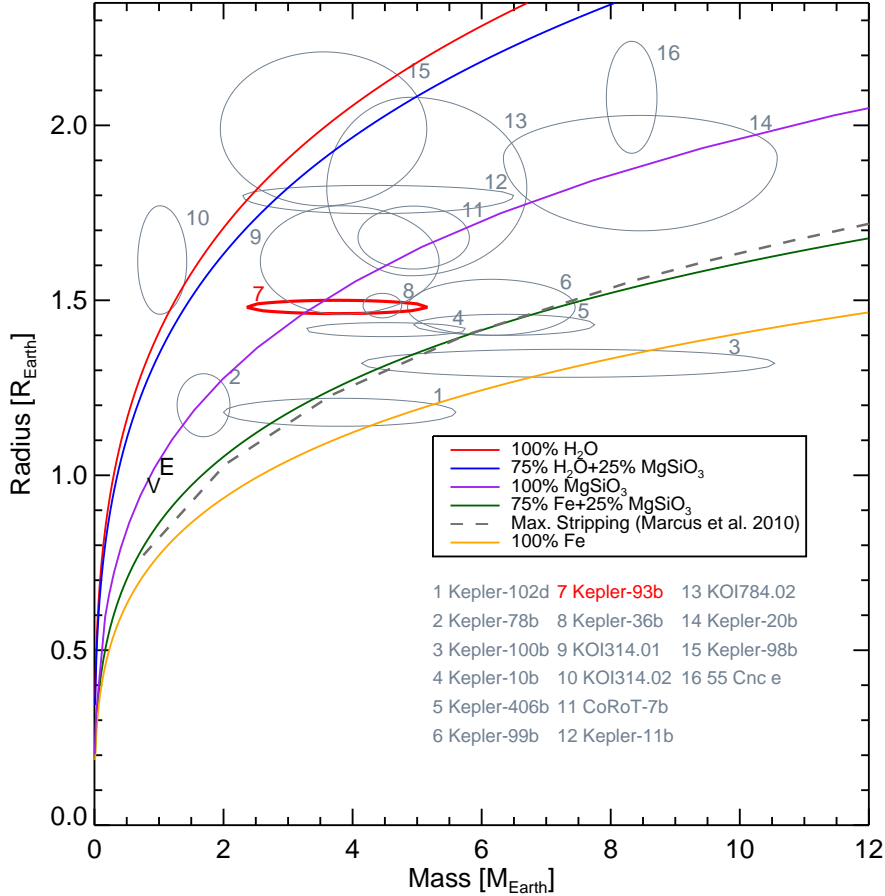} 
 \caption{The masses and radii for all known planets with radii
   $<$2$R_{\oplus}$ (within 1$\sigma$ uncertainty). We have included
   only planets with detected mass measurements of at least 2$\sigma$ certainty. We overplotted
   the theoretical mass-radius curves from \cite{Zeng13} for five
   cases (in order of increasing density): pure water, a mixture of
   75\% water and 25\% magnesium silicate, pure magnesium silicate, a mixture of 75\% iron and
   75\% magnesium silicate, and 100\% iron. We also include the theoretical maximum collisional stripping limit from \cite{Marcus10}, below which planets should not be able to acquire a larger iron fraction.} %The ephemeris e use to generate these O-C values is given in Table 2, and the individual transit times are given in Table %\ref{tbl:times}. The demarcation between long cadence and short cadence observations is shown with a dotted line.}
  \label{fig:mass_radius}
\end{center}
\end{figure}

\begin{figure}
\begin{center}
\includegraphics[width=4in]{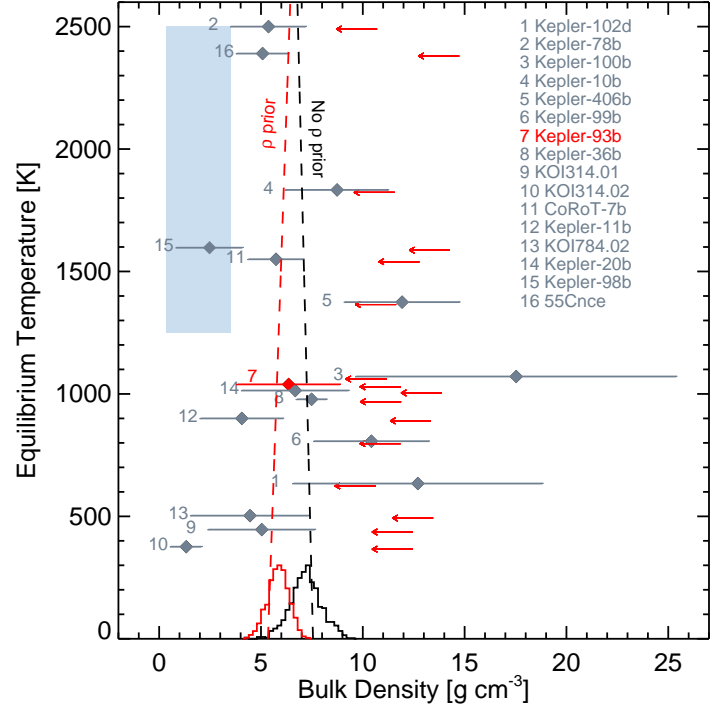}
\caption{Planetary bulk density as a function of equilibrium temperature, for known exoplanets smaller than 2.2 $R_{\oplus}$. Individual planets are labeled in order of their radius, from smallest to largest. These labels are the same as in Figure \ref{fig:mass_radius}. Red arrows immediately to the right of each error bar mark the upper limit on density from \cite{Marcus10}. Densities heavier than this value are not theoretically supportable, given the collisional-stripping model. Two dashed lines depict the best-fit relation between equilibrium temperature and density. The black line allows for unphysical densities, while the red line does not. The posterior distribution on the "mean" density of the sample is indicated on the bottom axis, for the same two cases. The shaded region corresponds to the parameter space in which \cite{Carter12} observed a dearth of planets.} 
\label{fig:teq_rho}
\end{center}
\end{figure}

In Figure \ref{fig:mass_radius}, we show the mass and radius measurements of the current sample of exoplanets smaller than 2.2 $R_{\oplus}$. The figure does not include the growing population of exoplanets with upper mass limits from radial velocity measurements or dynamical stability arguments.  We consider whether there exists evidence for a trend of bulk density with equilibrium temperature for this subset of planets. Their densities range from 1.3 g cm$^{-3}$ for KOI 314.02 \citep{Kipping14}, to 18 g cm$^{-3}$ for Kepler-100b \citep{Marcy14}. The exact cutoff of the transition between rocky and gaseous worlds likely resides between 1.5 and 2 $R_{\oplus}$, with both observations and theory informing our understanding (\citealt{Marcy14} and \citealt{Mordasini12}, respectively, among others). \cite{Rogers14} determined that the value at which half of planets are rocky and half gaseous occurs no higher than 1.6 $R_{\oplus}$. Whether or not equilibrium temperature also affects this relationship is what we test here. \cite{Carter12} studied the same parameter space for the known exoplanets with $M<10M_{\oplus}$. They noted a dearth of gaseous planets (densities less than 3.5 g cm$^{-3}$) hotter than 1250 K. They posited that evaporation plays a role in stripping the atmospheres from the planets with higher insolation. In Figure \ref{fig:teq_rho}, we overplot in gray the region where \cite{Carter12} observed a lack of planets. We note first that the  temperatures of exoplanets have large uncertainties. This is true even for worlds such as Kepler-78b, for which \cite{SanchisOjeda13} robustly detected the light emanating from the planet alone. For the purposes of this investigation, we assign the temperature reported by the authors in their discovery papers. We use the mean value within the range of possible temperatures, if the authors quote their uncertainty. If temperature was not among the quoted parameters, we use the equilibrium temperature value from the NASA Exoplanet Archive\footnote[1]{http://exoplanetarchive.ipac.caltech.edu/}. We assign the range in density uncertainty from the quoted mass and radius uncertainties. As we indicate in Figure \ref{fig:teq_rho}, we find a flat relationship between equilibrium temperature and density, indicating that temperature is not meaningfully predictive of the densities of planets within this radius range. However, this calculation presupposes no theoretical upper limit to density. Kepler-100b, for example, must constitute 70-100\% iron by mass to be consistent with its measured properties. \cite{Marcus10} derived an upper limit to iron mass fraction, assuming a collisional stripping history for denser worlds. This framework cannot lead to rocky planets with a fractional iron content above 75\% (this fraction varies slightly with radius).  We apply the hard prior that each planet cannot possess a density greater than the value computed by \cite{Marcus10} for this mass, and recompute the best-fit line. Now, we find that the best-fit line is one that predicts increasing density with stronger insolation, but the detection is indistinguishable from a flat line. The prior does meaningfully change the mean density of worlds between 1.0 and 1.5 $R_{\oplus}$. Taking their reported mass and radius measurements, the mean density of planets with 1.0$<R_{\oplus}<1.5$ is 10.0$\pm$1.5 g cm$^{-3}$, nearly twice as dense as Earth. This range is barely theoretical supportable, since worlds between 1.0 and 1.5 $R_{\oplus}$ have maximum theoretically plausible densities of 11 g cm$^{-3}$. We would conclude that Kepler-93b, with density of 6.3$\pm$ 2.6 g cm$^{-3}$, lies within the lightest 9\% of planets within 1.0$<R_{\oplus}<$1.5. Earth, with a density of 5.5 g cm$^{-3}$, resides 3$\sigma$ from the mean density of similarly sized worlds and ought to be rare indeed. In contrast, applying a prior of plausible density returns a mean value of 7.3$\pm$0.9 g cm$^{-3}$, and we would conclude that both Earth and Kepler-93b possess average density among similarly sized worlds.

\begin{figure}
\begin{center}
\includegraphics[width=4in]{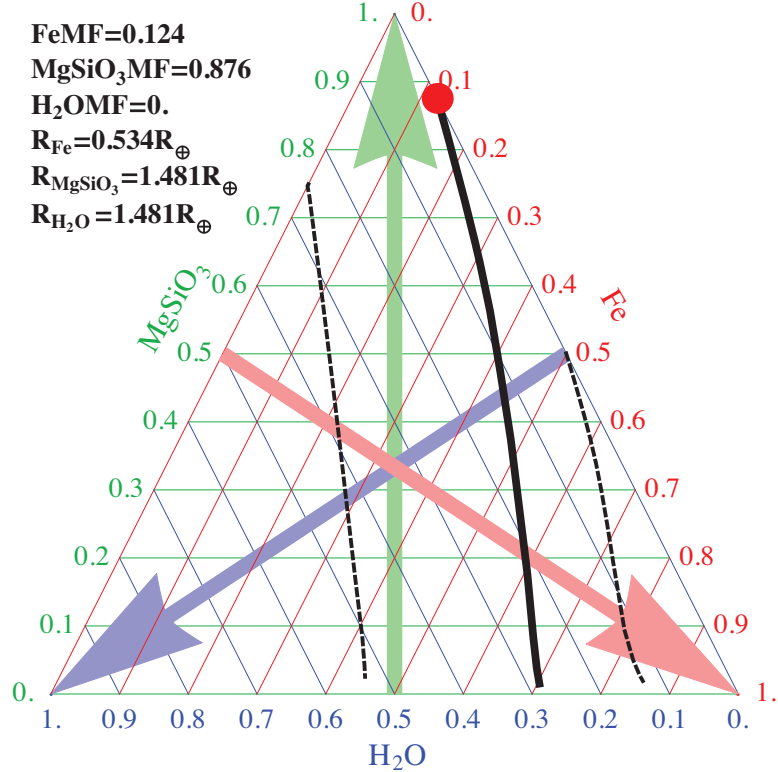}
\caption{Ternary diagram for Kepler-93b, using the models of \cite{Zeng13} comprised of water, magnesium silicate, and iron. The compositional degeneracy for the
best-fit value of mass and radius is depicted with a solid line, while the allowable
range for compositions within the radius and mass uncertainty is
indicated with dotted lines. The model with with no water is indicated with a red circle. The mass fraction and radial extent of each element (assuming complete differentiation) associated with this composition are indicated in upper left.} 
\label{fig:ternary}
\end{center}
\end{figure}

We compare the mass and radius of Kepler-93b against the theoretical compositions computed by \cite{Zeng13} for differing budgets of water, magnesium silicate, and iron. Figure \ref{fig:ternary} depicts the compositional ternary diagram for Kepler-93b, employing the theoretical models of \cite{Zeng13}. A 100\% magnesium silicate planet at the radius of Kepler-93b would have a mass of 3.3 $M_{\oplus}$, which is only 0.3$\sigma$ removed from the measured mass of $3.8\pm1.5$ $M_{\oplus}$. All mass fractions of magnesium silicate are plausible within the 1$\sigma$ error bar, though a pure water and iron planet is barely theoretically supportable. Within the 1$\sigma$ range in radius and mass that we report, iron mass fractions above 70\% are unphysical \citep{Marcus10}, and the entire remaining composition would need to be water for the planet to contain no rock. Given the age of the star of 6.6$\pm$0.9 Gyr, we consider it possible that all water content has been lost within Kepler-93b through outgassing. If this were the case, the planet would comprise 12.5\% iron by mass, and 87.5\% magnesium silicate.

\subsection{Likelihood of Extended Atmosphere}
To estimate of the likelihood of Kepler-93b's possessing an extended atmosphere, we first examine the criterion for Jean's escape for diatomic hydrogen. Given the temperature of 1037$\pm13$ K that we infer based upon the planet's insolation (and an assumption for its albedo of 0.3), the root-mean-squared velocity within the Maxwell-Boltzmann distribution for an $H_{2}$ molecule is 3600 m s$^{-1}$. In contrast, the escape speed from Kepler-93b is 18 000 m s$^{-1}$.  We assume that species with root-mean-square speeds higher than one sixth the escape speed will be lost from the atmosphere \cite{Seager10}, but we find $v_{\mbox{rms}/v_{\mbox{esc}}}\approx$5 for diatomic hydrogen, and so atmospheres of all kinds are theoretically plausible. We employ the metric created by
\cite{Kipping13} to assign a probability to the likelihood of an extended atmosphere, given the measured mass and radius posterior distributions. They provide a framework to calculate the posterior
on the quantity $R_{MAH}$, the physical extent of the planet's
atmosphere atop a minimally dense core of water as derived by
\cite{Zeng13}. They assume that the bulk composition of a super-Earth-sized planet cannot be made of a material lighter than water, and so if the radius of the planet is
significantly larger than inferred from this lower physical limit on
the density, it must possess an atmosphere to match the observed radius. We apply the radius for Kepler-93b
from Table \ref{tbl:res} of 1.478$^{+0.019}_{-0.019}$ $R_{\oplus}$ and
the mass of 3.8$\pm$1.5 $M_{\oplus}$ to find a probably of 3\% that the planet possesses an
extended atmosphere. Kepler-93b is therefore 97\% likely to be rocky in composition. In comparison, the published radius and mass posterior distributions for Kepler-36c and Kepler-10b return a
probability for an extended atmosphere of $<$0.01\% and 0.2\%,
respectively. These smaller probabilities, even though Kepler-36c and Kepler-10b are nearly the same size as Kepler-93b, are attributable to the slightly higher mass estimates for the former two worlds. We conclude that there exists a small but non-zero
probability that the planet possesses a significantly larger
atmosphere than the two planets most similar to it. 

\subsection{Secondary Eclipse Constraint}
\begin{figure}[h!]
\begin{center}
 \includegraphics[width=5in]{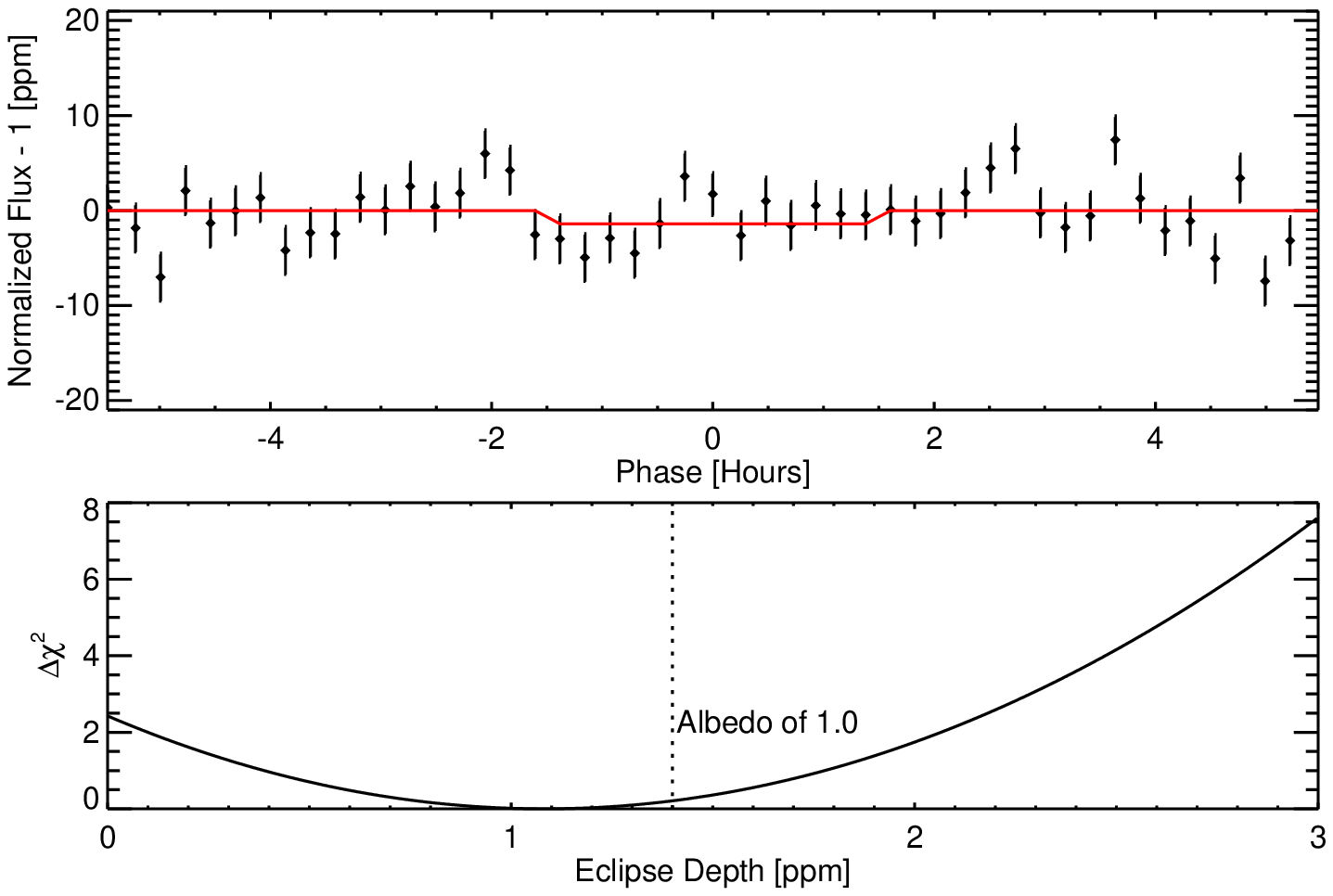} 
 \caption{{\it Top panel:} Light curve of Kepler-93b, centered on an
   orbital phase of 0.5. We have overplotted the eclipse model
   corresponding to the maximum albedo of 1.0. {\it Bottom panel:} The
   $\chi^{2}$ improvement afforded by eclipse models, as a function of
   the depth of the secondary eclipse. We find only an upper limit to the eclipse depth: it must be $<$2.5 ppm at a phase of 0.5, with 2$\sigma$ confidence. This is not physically constraining, since the predicted depth produced by a planet with albedo of 1.0 is 1.4 ppm.}
  \label{fig:eclipse}
\end{center}
\end{figure}

Assuming again the Bond albedo ($A_{B}$)of 0.3, the expected relative depth due to
reflected light is given by $\delta_{ref}=A_{B}(R_{p}/a)^{2}$. This
value is 0.4 ppm for $A=0.3$, but could be as high as 1.4 ppm for an albedo of
1.0. The expected secondary eclipse depth due to the emitted light of the
planet is given by $\delta_{em}=(R_{p}/R_{\star})^{2}\cdot
B_{\lambda}(T_{p})/B_{\lambda}(T_{\star})$. Assuming again an albedo
of 0.3, wavelength of 700 nm (in the middle of the \kepler\ bandpass)
to estimate $B_{\lambda}(T)$, $\delta_{em}$ is of order $10^{-10}$ and
so contributes negligibly to the expected eclipse depth. In Figure \ref{fig:eclipse}, we show the phased light
curve, binned in increments of 12 minutes, and centered on a phase of 0.5. We have treated these data similarly to the data in-transit, by fitting a linear baseline with time to the observations immediately adjacent to each predicted eclipse and dividing this line from the eclipse observations. The scatter per 15-minute error bar is 2.4 ppm, so for a 173 minute secondary eclipse, the predicted depth error bar at $1\sigma$ is 0.7 ppm. All physical depths (that is, $<1.4$ ppm) should therefore furnish $\chi^{2}$ fits within 2$\sigma$ of one another. We conclude that the data cannot meaningfully distinguish between eclipse depths within the physical range of albedos.  Only eclipses deeper than 6 ppm are ruled
out at 3$\sigma$ significance, and this value is too high to be physically meaningful. We repeated the eclipse search in intervals of 1 hr from phases of 0.4 to 0.6 of the orbital phase (corresponding to 11 hours preceding and following the eclipse time at 0.5 the orbital phase). We produce a new eclipse light curve to test each putative eclipse time, by fitting the baseline to out-of-eclipse observations adjacent to the new ephemeris (to avoid normalizing out any authentic eclipse signal). We report no statistically significant detection of an eclipse signal over this range, and set an upper limit on the eclipse depth of 2.5 ppm with 2$\sigma$ confidence.

\section{Conclusions}
We have presented the characterization of Kepler-93b, a rocky planet orbiting one of the brightest stars studied by \kepler.  

\begin{itemize}
\item We measure the asteroseismic spectrum of Kepler-93. With a mass of $0.911 \pm 0.033$ $M_{\odot}$, it is one of the lowest mass subjects of asteroseismic study. We measure its radius to be $0.919 \pm 0.011$ $R_{\odot}$ and its mean density to be 1.652 $\pm0.006$ g cm$^{-3}$.
\item We measure the radius of Kepler-93b to be 1.481$\pm0.019$ $R_{\oplus}$. Kepler-93b is the most precisely measured planet outside of the solar system, and among the only four exoplanets whose sizes are known to within 0.03 $R_{\oplus}$. Applying our knowledge of the star from asteroseimology to fit the transit light curve reduced the uncertainty on the planet radius by a third.
\item We measure a consistent transit depth for Kepler-93b with both \kepler\ and \spitzer\ observations. We used Kepler-93b as a test subject for the peak-up observing mode with \spitzer. We find that peak-up reduces the uncertainty with which we measure the planet-to-star radius ratio by a factor of 2.
\item From the radial velocity measurements of the mass of Kepler-93b, we conclude that it possesses a mean density of 6.3$\pm$2.6 g cm$^{-3}$. It likely has a terrestrial composition. Applying a theoretical density prior on the known exoplanets smaller than 1.5 $R_{\oplus}$, we find that the average density of planets in this range is 7.3$\pm$0.9 g cm$^{-3}$ (this density is nearly unphysically high without the prior). We conclude that Kepler-93b has a density that is average among similarly-sized planets.
\end{itemize}

The best way for increasing our knowledge about the planet's nature is by continued radial velocity observations of the star. Firstly, an extended baseline of radial velocity measurements will constrain the nature of Kepler-93c, the long-period companion. The current four year baseline only proves that this object has an orbital period longer than 6 years, and that it must be larger than 3 Jupiter masses. The story of the formation of Kepler-93b and its migration to its current highly irradiated location will be illuminated in part by an understanding of this companion's mass and eccentricity. And secondly, the continued monitoring of the shorter-term radial velocity signature of Kepler-93b will allow us to winnow down the allowable parameter space of its composition. We conduct a sample calculation of the likelihood of an extended atmosphere on Kepler-93b with a refined mass estimate of the planet. If the uncertainty was reduced by a factor of two from its current value of 1.5 $M_{\oplus}$ to 0.75 $M_{\oplus}$, and the most probable value were refined to 1.5$\sigma$ below the current most probable value of 3.8 $M_{\oplus}$, then the likelihood of an extended atmosphere is upwardly revised to 25\%. The asteroseismic study of this world has returned a strong constraint of 6.6$\pm$0.9 Gyr on the planet's age. If it is plausible for the water on the planet to be outgassed and lost under its insolation conditions on this timescale, we could reconsider the magnesium and iron content of its core. If the planet is made of only the latter two materials, models indicate that its iron mass fraction would be only 40\% that of the Earth's, which is 32.5\% iron by mass \citep{Morgan80}. The unprecedented determination of the radius of Kepler-93b to 120 km speaks to the increasing precision with which we are able to examine planets orbiting other stars. Exoplanetary systems such as Kepler-93 that can support simultaneous studies of stellar structure, exoplanetary radius, and exoplanetary mass comprise valuable laboratories for future study.

\acknowledgements

This work was performed in part under contract with the California Institute of
Technology (Caltech) funded by NASA through the Sagan Fellowship
Program. It was conducted with observations made with the \spitzer\
Space Telescope, which is operated by the Jet Propulsion Laboratory,
California Institute of Technology under a contract with NASA. Support
for this work was provided by NASA through an award issued by
JPL/Caltech. We thank the \spitzer\ team at the Infrared Processing and Analysis
Center in Pasadena, California, and in particular Nancy Silbermann for
scheduling the \spitzer\ observations of this program. This work is also based on observations made with
\kepler, which was competitively selected as the tenth Discovery
mission. Funding for this mission is provided by NASA's Science
Mission Directorate. The authors would like to thank the many people
who generously gave so much their time to make this Mission a success. This research has made use of the NASA Exoplanet Archive, which is operated by the California Institute of Technology, under contract with the National Aeronautics and Space Administration under the Exoplanet Exploration Program. S. Ballard thanks Geoffrey Marcy for helpful discussions
about the radial velocity signature of Kepler-93. 
We acknowledge support through \kepler\ Participatory Science
Awards NNX12AC77G and NNX09AB53G, awarded to D.C. This publication was made possible in part through the support of a grant from the John Templeton Foundation. The opinions expressed in this publication are those of the authors and do not necessarily reflect the views of the John Templeton Foundation. 
W.J.C., T.L.C., G.R.D., Y.E. and A.M. acknowledge the support of the
UK Science and Technology Facilities Council (STFC). S. Basu
acknowledges support from NSF grant AST-1105930 and NASA grant NNX13AE70G. Funding for the Stellar Astrophysics Centre is provided by The Danish National Research Foundation (Grant agreement no.: DNRF106). The research is supported by the ASTERISK project (ASTERoseismic Investigations with SONG and Kepler) funded by the European Research Council (Grant agreement no.: 267864).  S.H. acknowledges financial support from the Netherlands Organisation for Scientific Research (NWO). The research leading to these results has received funding from the European Research Council under the European Community's Seventh Framework Programme (FP7/2007-2013) / ERC grant agreement no 338251 (StellarAges). T.S.M. acknowledges NASA grant NNX13AE91G. D.S. is supported by the Australian Research Council. DH acknowledges support by an appointment to the NASA Postdoctoral Program at Ames Research Center administered by Oak Ridge Associated Universities, and NASA Grant NNX14AB92G issued through the Kepler Participating Scientist Program. Computational time on Kraken at the National Institute of Computational Sciences was provided through NSF TeraGrid allocation TG-AST090107. We are also grateful for support from the International Space Science Institute (ISSI).

\newpage
\bibliographystyle{apj} 
%\bibliography{allrefs}

\begin{thebibliography}{108}
\expandafter\ifx\csname natexlab\endcsname\relax\def\natexlab#1{#1}\fi

\bibitem[{{Angulo} {et~al.}(1999){Angulo}, {Arnould}, {Rayet}, {Descouvemont},
  {Baye}, {Leclercq-Willain}, {Coc}, {Barhoumi}, {Aguer}, {Rolfs}, {Kunz},
  {Hammer}, {Mayer}, {Paradellis}, {Kossionides}, {Chronidou}, {Spyrou},
  {degl'Innocenti}, {Fiorentini}, {Ricci}, {Zavatarelli}, {Providencia},
  {Wolters}, {Soares}, {Grama}, {Rahighi}, {Shotter}, \& {Lamehi
  Rachti}}]{Angulo99}
{Angulo}, C., {et~al.} 1999, Nuclear Physics A, 656, 3

\bibitem[{{Appourchaux}(2011)}]{Appourchaux11}
{Appourchaux}, T. 2011, in: Asteroseismology, ed. P Pall\'e, vol. 23 of Canary
  Islands Winter School of Astrophysics, Cambridge University Press, in press
  (arXiv:1103.5352)

\bibitem[{{Argabright} {et~al.}(2008){Argabright}, {VanCleve}, {Bachtell},
  {Hegge}, {McArthur}, {Dumont}, {Rudeen}, {Pullen}, {Teusch}, {Tennant}, \&
  {Atcheson}}]{Argabright08}
{Argabright}, V.~S., {et~al.} 2008, in Society of Photo-Optical Instrumentation
  Engineers (SPIE) Conference Series, Vol. 7010, Society of Photo-Optical
  Instrumentation Engineers (SPIE) Conference Series

\bibitem[{{Ballard} {et~al.}(2010){Ballard}, {Charbonneau}, {Deming},
  {Knutson}, {Christiansen}, {Holman}, {Fabrycky}, {Seager}, \&
  {A'Hearn}}]{Ballard10b}
{Ballard}, S., {et~al.} 2010, \pasp, 122, 1341

\bibitem[{{Ballard} {et~al.}(2013){Ballard}, {Charbonneau}, {Fressin},
  {Torres}, {Irwin}, {Desert}, {Newton}, {Mann}, {Ciardi}, {Crepp}, {Henze},
  {Bryson}, {Howell}, {Horch}, {Everett}, \& {Shporer}}]{Ballard13}
---. 2013, \apj, 773, 98

\bibitem[{{Ballard} {et~al.}(2011){Ballard}, {Fabrycky}, {Fressin},
  {Charbonneau}, {Desert}, {Torres}, {Marcy}, {Burke}, {Isaacson}, {Henze},
  {Steffen}, {Ciardi}, {Howell}, {Cochran}, {Endl}, {Bryson}, {Rowe}, {Holman},
  {Lissauer}, {Jenkins}, {Still}, {Ford}, {Christiansen}, {Middour}, {Haas},
  {Li}, {Hall}, {McCauliff}, {Batalha}, {Koch}, \& {Borucki}}]{Ballard11b}
---. 2011, \apj, 743, 200

\bibitem[{{Barclay} {et~al.}(2013){Barclay}, {Rowe}, {Lissauer}, {Huber},
  {Fressin}, {Howell}, {Bryson}, {Chaplin}, {D{\'e}sert}, {Lopez}, {Marcy},
  {Mullally}, {Ragozzine}, {Torres}, {Adams}, {Agol}, {Barrado}, {Basu},
  {Bedding}, {Buchhave}, {Charbonneau}, {Christiansen},
  {Christensen-Dalsgaard}, {Ciardi}, {Cochran}, {Dupree}, {Elsworth},
  {Everett}, {Fischer}, {Ford}, {Fortney}, {Geary}, {Haas}, {Handberg},
  {Hekker}, {Henze}, {Horch}, {Howard}, {Hunter}, {Isaacson}, {Jenkins},
  {Karoff}, {Kawaler}, {Kjeldsen}, {Klaus}, {Latham}, {Li}, {Lillo-Box},
  {Lund}, {Lundkvist}, {Metcalfe}, {Miglio}, {Morris}, {Quintana}, {Stello},
  {Smith}, {Still}, \& {Thompson}}]{Barclay13}
{Barclay}, T., {et~al.} 2013, \nat, 494, 452

\bibitem[{{Batalha} {et~al.}(2011){Batalha}, {Borucki}, {Bryson}, {Buchhave},
  {Caldwell}, {Christensen-Dalsgaard}, {Ciardi}, {Dunham}, {Fressin},
  {Gautier}, {Gilliland}, {Haas}, {Howell}, {Jenkins}, {Kjeldsen}, {Koch},
  {Latham}, {Lissauer}, {Marcy}, {Rowe}, {Sasselov}, {Seager}, {Steffen},
  {Torres}, {Basri}, {Brown}, {Charbonneau}, {Christiansen}, {Clarke},
  {Cochran}, {Dupree}, {Fabrycky}, {Fischer}, {Ford}, {Fortney}, {Girouard},
  {Holman}, {Johnson}, {Isaacson}, {Klaus}, {Machalek}, {Moorehead},
  {Morehead}, {Ragozzine}, {Tenenbaum}, {Twicken}, {Quinn}, {VanCleve},
  {Walkowicz}, {Welsh}, {Devore}, \& {Gould}}]{Batalha11}
{Batalha}, N.~M., {et~al.} 2011, \apj, 729, 27

\bibitem[{{Batalha} {et~al.}(2010){Batalha}, {Borucki}, {Koch}, {Bryson},
  {Haas}, {Brown}, {Caldwell}, {Hall}, {Gilliland}, {Latham}, {Meibom}, \&
  {Monet}}]{Batalha10}
---. 2010, \apjl, 713, L109

\bibitem[{Borucki {et~al.}(2013)Borucki, Agol, Fressin, Kaltenegger, Rowe,
  Isaacson, Fischer, Batalha, Lissauer, Marcy, Fabrycky, D{\'e}sert, Bryson,
  Barclay, Bastien, Boss, Brugamyer, Buchhave, Burke, Caldwell, Carter,
  Charbonneau, Crepp, Christensen-Dalsgaard, Christiansen, Ciardi, Cochran,
  DeVore, Doyle, Dupree, Endl, Everett, Ford, Fortney, Gautier, Geary, Gould,
  Haas, Henze, Howard, Howell, Huber, Jenkins, Kjeldsen, Kolbl, Kolodziejczak,
  Latham, Lee, Lopez, Mullally, Orosz, Prsa, Quintana, Sasselov, Seader,
  Shporer, Steffen, Still, Tenenbaum, Thompson, Torres, Twicken, Welsh, \&
  Winn}]{Borucki13}
Borucki, W.~J., {et~al.} 2013, Science

\bibitem[{{Borucki} {et~al.}(2011){Borucki}, {Koch}, {Basri}, {Batalha},
  {Brown}, {Bryson}, {Caldwell}, {Christensen-Dalsgaard}, {Cochran}, {DeVore},
  {Dunham}, {Gautier}, {Geary}, {Gilliland}, {Gould}, {Howell}, {Jenkins},
  {Latham}, {Lissauer}, {Marcy}, {Rowe}, {Sasselov}, {Boss}, {Charbonneau},
  {Ciardi}, {Doyle}, {Dupree}, {Ford}, {Fortney}, {Holman}, {Seager},
  {Steffen}, {Tarter}, {Welsh}, {Allen}, {Buchhave}, {Christiansen}, {Clarke},
  {Das}, {D{\'e}sert}, {Endl}, {Fabrycky}, {Fressin}, {Haas}, {Horch},
  {Howard}, {Isaacson}, {Kjeldsen}, {Kolodziejczak}, {Kulesa}, {Li}, {Lucas},
  {Machalek}, {McCarthy}, {MacQueen}, {Meibom}, {Miquel}, {Prsa}, {Quinn},
  {Quintana}, {Ragozzine}, {Sherry}, {Shporer}, {Tenenbaum}, {Torres},
  {Twicken}, {Van Cleve}, {Walkowicz}, {Witteborn}, \& {Still}}]{Borucki11}
{Borucki}, W.~J., {et~al.} 2011, \apj, 736, 19

\bibitem[{{Borucki} {et~al.}(2012){Borucki}, {Koch}, {Batalha}, {Bryson},
  {Rowe}, {Fressin}, {Torres}, {Caldwell}, {Christensen-Dalsgaard}, {Cochran},
  {DeVore}, {Gautier}, {Geary}, {Gilliland}, {Gould}, {Howell}, {Jenkins},
  {Latham}, {Lissauer}, {Marcy}, {Sasselov}, {Boss}, {Charbonneau}, {Ciardi},
  {Kaltenegger}, {Doyle}, {Dupree}, {Ford}, {Fortney}, {Holman}, {Steffen},
  {Mullally}, {Still}, {Tarter}, {Ballard}, {Buchhave}, {Carter},
  {Christiansen}, {Demory}, {D{\'e}sert}, {Dressing}, {Endl}, {Fabrycky},
  {Fischer}, {Haas}, {Henze}, {Horch}, {Howard}, {Isaacson}, {Kjeldsen},
  {Johnson}, {Klaus}, {Kolodziejczak}, {Barclay}, {Li}, {Meibom}, {Prsa},
  {Quinn}, {Quintana}, {Robertson}, {Sherry}, {Shporer}, {Tenenbaum},
  {Thompson}, {Twicken}, {Van Cleve}, {Welsh}, {Basu}, {Chaplin}, {Miglio},
  {Kawaler}, {Arentoft}, {Stello}, {Metcalfe}, {Verner}, {Karoff}, {Lundkvist},
  {Lund}, {Handberg}, {Elsworth}, {Hekker}, {Huber}, {Bedding}, \&
  {Rapin}}]{Borucki12}
---. 2012, \apj, 745, 120

\bibitem[{{Brown} \& {Gilliland}(1994)}]{Brown94}
{Brown}, T.~M., \& {Gilliland}, R.~L. 1994, \araa, 32, 37

\bibitem[{{Bruntt} {et~al.}(2012){Bruntt}, {Basu}, {Smalley}, {Chaplin},
  {Verner}, {Bedding}, {Catala}, {Gazzano}, {Molenda-{\.Z}akowicz}, {Thygesen},
  {Uytterhoeven}, {Hekker}, {Huber}, {Karoff}, {Mathur}, {Mosser},
  {Appourchaux}, {Campante}, {Elsworth}, {Garc{\'{\i}}a}, {Handberg},
  {Metcalfe}, {Quirion}, {R{\'e}gulo}, {Roxburgh}, {Stello},
  {Christensen-Dalsgaard}, {Kawaler}, {Kjeldsen}, {Morris}, {Quintana}, \&
  {Sanderfer}}]{Bruntt12}
{Bruntt}, H., {et~al.} 2012, \mnras, 423, 122

\bibitem[{{Buchhave} {et~al.}(2011){Buchhave}, {Latham}, {Carter},
  {D{\'e}sert}, {Torres}, {Adams}, {Bryson}, {Charbonneau}, {Ciardi}, {Kulesa},
  {Dupree}, {Fischer}, {Fressin}, {Gautier}, {Gilliland}, {Howell}, {Isaacson},
  {Jenkins}, {Marcy}, {McCarthy}, {Rowe}, {Batalha}, {Borucki}, {Brown},
  {Caldwell}, {Christiansen}, {Cochran}, {Deming}, {Dunham}, {Everett}, {Ford},
  {Fortney}, {Geary}, {Girouard}, {Haas}, {Holman}, {Horch}, {Klaus},
  {Knutson}, {Koch}, {Kolodziejczak}, {Lissauer}, {Machalek}, {Mullally},
  {Still}, {Quinn}, {Seager}, {Thompson}, \& {Van Cleve}}]{Buchave11}
{Buchhave}, L.~A., {et~al.} 2011, \apjs, 197, 3

\bibitem[{{Buchhave} {et~al.}(2012){Buchhave}, {Latham}, {Johansen},
  {Bizzarro}, {Torres}, {Rowe}, {Batalha}, {Borucki}, {Brugamyer}, {Caldwell},
  {Bryson}, {Ciardi}, {Cochran}, {Endl}, {Esquerdo}, {Ford}, {Geary},
  {Gilliland}, {Hansen}, {Isaacson}, {Laird}, {Lucas}, {Marcy}, {Morse},
  {Robertson}, {Shporer}, {Stefanik}, {Still}, \& {Quinn}}]{Buchave12}
---. 2012, \nat, 486, 375

\bibitem[{{Caldwell} {et~al.}(2010){Caldwell}, {van Cleve}, {Jenkins},
  {Argabright}, {Kolodziejczak}, {Dunham}, {Geary}, {Tenenbaum},
  {Chandrasekaran}, {Li}, {Wu}, \& {von Wilpert}}]{Caldwell10}
{Caldwell}, D.~A., {et~al.} 2010, in Society of Photo-Optical Instrumentation
  Engineers (SPIE) Conference Series, Vol. 7731, Society of Photo-Optical
  Instrumentation Engineers (SPIE) Conference Series

\bibitem[{{Carter} {et~al.}(2012){Carter}, {Agol}, {Chaplin}, {Basu},
  {Bedding}, {Buchhave}, {Christensen-Dalsgaard}, {Deck}, {Elsworth},
  {Fabrycky}, {Ford}, {Fortney}, {Hale}, {Handberg}, {Hekker}, {Holman},
  {Huber}, {Karoff}, {Kawaler}, {Kjeldsen}, {Lissauer}, {Lopez}, {Lund},
  {Lundkvist}, {Metcalfe}, {Miglio}, {Rogers}, {Stello}, {Borucki}, {Bryson},
  {Christiansen}, {Cochran}, {Geary}, {Gilliland}, {Haas}, {Hall}, {Howard},
  {Jenkins}, {Klaus}, {Koch}, {Latham}, {MacQueen}, {Sasselov}, {Steffen},
  {Twicken}, \& {Winn}}]{Carter12}
{Carter}, J.~A., {et~al.} 2012, Science, 337, 556

\bibitem[{{Carter} \& {Winn}(2009)}]{Carter09}
{Carter}, J.~A., \& {Winn}, J.~N. 2009, \apj, 704, 51

\bibitem[{{Chaplin} {et~al.}(2011{\natexlab{a}}){Chaplin}, {Kjeldsen},
  {Bedding}, {Christensen-Dalsgaard}, {Gilliland}, {Kawaler}, {Appourchaux},
  {Elsworth}, {Garc{\'{\i}}a}, {Houdek}, {Karoff}, {Metcalfe},
  {Molenda-{\.Z}akowicz}, {Monteiro}, {Thompson}, {Verner}, {Batalha},
  {Borucki}, {Brown}, {Bryson}, {Christiansen}, {Clarke}, {Jenkins}, {Klaus},
  {Koch}, {An}, {Ballot}, {Basu}, {Benomar}, {Bonanno}, {Broomhall},
  {Campante}, {Corsaro}, {Creevey}, {Esch}, {Gai}, {Gaulme}, {Hale},
  {Handberg}, {Hekker}, {Huber}, {Mathur}, {Mosser}, {New}, {Pinsonneault},
  {Pricopi}, {Quirion}, {R{\'e}gulo}, {Roxburgh}, {Salabert}, {Stello}, \&
  {Suran}}]{Chaplin11b}
{Chaplin}, W.~J., {et~al.} 2011{\natexlab{a}}, \apj, 732, 54

\bibitem[{{Chaplin} {et~al.}(2011{\natexlab{b}}){Chaplin}, {Kjeldsen},
  {Christensen-Dalsgaard}, {Basu}, {Miglio}, {Appourchaux}, {Bedding},
  {Elsworth}, {Garc{\'{\i}}a}, {Gilliland}, {Girardi}, {Houdek}, {Karoff},
  {Kawaler}, {Metcalfe}, {Molenda-{\.Z}akowicz}, {Monteiro}, {Thompson},
  {Verner}, {Ballot}, {Bonanno}, {Brand{\~a}o}, {Broomhall}, {Bruntt},
  {Campante}, {Corsaro}, {Creevey}, {Do{\u g}an}, {Esch}, {Gai}, {Gaulme},
  {Hale}, {Handberg}, {Hekker}, {Huber}, {Jim{\'e}nez}, {Mathur}, {Mazumdar},
  {Mosser}, {New}, {Pinsonneault}, {Pricopi}, {Quirion}, {R{\'e}gulo},
  {Salabert}, {Serenelli}, {Silva Aguirre}, {Sousa}, {Stello}, {Stevens},
  {Suran}, {Uytterhoeven}, {White}, {Borucki}, {Brown}, {Jenkins}, {Kinemuchi},
  {Van Cleve}, \& {Klaus}}]{Chaplin11a}
---. 2011{\natexlab{b}}, Science, 332, 213

\bibitem[{{Chaplin} {et~al.}(2013){Chaplin}, {Sanchis-Ojeda}, {Campante},
  {Handberg}, {Stello}, {Winn}, {Basu}, {Christensen-Dalsgaard}, {Davies},
  {Metcalfe}, {Buchhave}, {Fischer}, {Bedding}, {Cochran}, {Elsworth},
  {Gilliland}, {Hekker}, {Huber}, {Isaacson}, {Karoff}, {Kawaler}, {Kjeldsen},
  {Latham}, {Lund}, {Lundkvist}, {Marcy}, {Miglio}, {Barclay}, \&
  {Lissauer}}]{Chaplin13}
---. 2013, \apj, 766, 101

\bibitem[{{Charbonneau} {et~al.}(2005){Charbonneau}, {Allen}, {Megeath},
  {Torres}, {Alonso}, {Brown}, {Gilliland}, {Latham}, {Mandushev}, {O'Donovan},
  \& {Sozzetti}}]{Charbonneau05}
{Charbonneau}, D., {et~al.} 2005, \apj, 626, 523

\bibitem[{{Christensen-Dalsgaard}(2008{\natexlab{a}})}]{Christensen08b}
{Christensen-Dalsgaard}, J. 2008{\natexlab{a}}, \apss, 316, 113

\bibitem[{{Christensen-Dalsgaard}(2008{\natexlab{b}})}]{Christensen08a}
---. 2008{\natexlab{b}}, \apss, 316, 13

\bibitem[{{Cochran} {et~al.}(2011){Cochran}, {Fabrycky}, {Torres}, {Fressin},
  {D{\'e}sert}, {Ragozzine}, {Sasselov}, {Fortney}, {Rowe}, {Brugamyer},
  {Bryson}, {Carter}, {Ciardi}, {Howell}, {Steffen}, {Borucki}, {Koch}, {Winn},
  {Welsh}, {Uddin}, {Tenenbaum}, {Still}, {Seager}, {Quinn}, {Mullally},
  {Miller}, {Marcy}, {MacQueen}, {Lucas}, {Lissauer}, {Latham}, {Knutson},
  {Kinemuchi}, {Johnson}, {Jenkins}, {Isaacson}, {Howard}, {Horch}, {Holman},
  {Henze}, {Haas}, {Gilliland}, {Gautier}, {Ford}, {Fischer}, {Everett},
  {Endl}, {Demory}, {Deming}, {Charbonneau}, {Caldwell}, {Buchhave}, {Brown},
  \& {Batalha}}]{Cochran11}
{Cochran}, W.~D., {et~al.} 2011, \apjs, 197, 7

\bibitem[{{Demarque} {et~al.}(2008){Demarque}, {Guenther}, {Li}, {Mazumdar}, \&
  {Straka}}]{Demarque08}
{Demarque}, P., {Guenther}, D.~B., {Li}, L.~H., {Mazumdar}, A., \& {Straka},
  C.~W. 2008, \apss, 316, 31

\bibitem[{{Demory} {et~al.}(2011){Demory}, {Gillon}, {Deming}, {Valencia},
  {Seager}, {Benneke}, {Lovis}, {Cubillos}, {Harrington}, {Stevenson}, {Mayor},
  {Pepe}, {Queloz}, {S{\'e}gransan}, \& {Udry}}]{Demory11b}
{Demory}, B.-O., {et~al.} 2011, \aap, 533, A114

\bibitem[{{D{\'e}sert} {et~al.}(2011){D{\'e}sert}, {Bean}, {Miller-Ricci
  Kempton}, {Berta}, {Charbonneau}, {Irwin}, {Fortney}, {Burke}, \&
  {Nutzman}}]{Desert11}
{D{\'e}sert}, J.-M., {et~al.} 2011, \apjl, 731, L40

\bibitem[{{Dotter} {et~al.}(2008){Dotter}, {Chaboyer}, {Jevremovi{\'c}},
  {Kostov}, {Baron}, \& {Ferguson}}]{Dotter08}
{Dotter}, A., {Chaboyer}, B., {Jevremovi{\'c}}, D., {Kostov}, V., {Baron}, E.,
  \& {Ferguson}, J.~W. 2008, \apjs, 178, 89

\bibitem[{{Doyle} {et~al.}(2011){Doyle}, {Carter}, {Fabrycky}, {Slawson},
  {Howell}, {Winn}, {Orosz}, {Prsa}, {Welsh}, {Quinn}, {Latham}, {Torres},
  {Buchhave}, {Marcy}, {Fortney}, {Shporer}, {Ford}, {Lissauer}, {Ragozzine},
  {Rucker}, {Batalha}, {Jenkins}, {Borucki}, {Koch}, {Middour}, {Hall},
  {McCauliff}, {Fanelli}, {Quintana}, {Holman}, {Caldwell}, {Still},
  {Stefanik}, {Brown}, {Esquerdo}, {Tang}, {Furesz}, {Geary}, {Berlind},
  {Calkins}, {Short}, {Steffen}, {Sasselov}, {Dunham}, {Cochran}, {Boss},
  {Haas}, {Buzasi}, \& {Fischer}}]{Doyle11}
{Doyle}, L.~R., {et~al.} 2011, Science, 333, 1602

\bibitem[{{Dressing} \& {Charbonneau}(2013)}]{Dressing13}
{Dressing}, C.~D., \& {Charbonneau}, D. 2013, \apj, 767, 95

\bibitem[{{Fazio} {et~al.}(2004){Fazio}, {Hora}, {Allen}, {Ashby}, {Barmby},
  {Deutsch}, {Huang}, {Kleiner}, {Marengo}, {Megeath}, {Melnick}, {Pahre},
  {Patten}, {Polizotti}, {Smith}, {Taylor}, {Wang}, {Willner}, {Hoffmann},
  {Pipher}, {Forrest}, {McMurty}, {McCreight}, {McKelvey}, {McMurray}, {Koch},
  {Moseley}, {Arendt}, {Mentzell}, {Marx}, {Losch}, {Mayman}, {Eichhorn},
  {Krebs}, {Jhabvala}, {Gezari}, {Fixsen}, {Flores}, {Shakoorzadeh}, {Jungo},
  {Hakun}, {Workman}, {Karpati}, {Kichak}, {Whitley}, {Mann}, {Tollestrup},
  {Eisenhardt}, {Stern}, {Gorjian}, {Bhattacharya}, {Carey}, {Nelson},
  {Glaccum}, {Lacy}, {Lowrance}, {Laine}, {Reach}, {Stauffer}, {Surace},
  {Wilson}, {Wright}, {Hoffman}, {Domingo}, \& {Cohen}}]{Fazio04}
{Fazio}, G.~G., {et~al.} 2004, \apjs, 154, 10

\bibitem[{{Ferguson} {et~al.}(2005){Ferguson}, {Alexander}, {Allard}, {Barman},
  {Bodnarik}, {Hauschildt}, {Heffner-Wong}, \& {Tamanai}}]{Ferguson05}
{Ferguson}, J.~W., {Alexander}, D.~R., {Allard}, F., {Barman}, T., {Bodnarik},
  J.~G., {Hauschildt}, P.~H., {Heffner-Wong}, A., \& {Tamanai}, A. 2005, \apj,
  623, 585

\bibitem[{{Fogtmann-Schulz} {et~al.}(2014){Fogtmann-Schulz}, {Hinrup}, {Van
  Eylen}, {Christensen-Dalsgaard}, {Kjeldsen}, {Silva Aguirre}, \&
  {Tingley}}]{Fogtmann14}
{Fogtmann-Schulz}, A., {Hinrup}, B., {Van Eylen}, V., {Christensen-Dalsgaard},
  J., {Kjeldsen}, H., {Silva Aguirre}, V., \& {Tingley}, B. 2014, \apj, 781, 67

\bibitem[{{Ford}(2005)}]{Ford05}
{Ford}, E.~B. 2005, \aj, 129, 1706

\bibitem[{{Formicola} {et~al.}(2004){Formicola}, {Imbriani}, {Costantini},
  {Angulo}, {Bemmerer}, {Bonetti}, {Broggini}, {Corvisiero}, {Cruz},
  {Descouvemont}, {F{\"u}l{\"o}p}, {Gervino}, {Guglielmetti}, {Gustavino},
  {Gy{\"u}rky}, {Jesus}, {Junker}, {Lemut}, {Menegazzo}, {Prati}, {Roca},
  {Rolfs}, {Romano}, {Rossi Alvarez}, {Sch{\"u}mann}, {Somorjai}, {Straniero},
  {Strieder}, {Terrasi}, {Trautvetter}, {Vomiero}, \&
  {Zavatarelli}}]{Formicola04}
{Formicola}, A., {et~al.} 2004, Physics Letters B, 591, 61

\bibitem[{{Fressin} {et~al.}(2013){Fressin}, {Torres}, {Charbonneau}, {Bryson},
  {Christiansen}, {Dressing}, {Jenkins}, {Walkowicz}, \& {Batalha}}]{Fressin13}
{Fressin}, F., {et~al.} 2013, \apj, 766, 81

\bibitem[{{Fressin} {et~al.}(2011){Fressin}, {Torres}, {D{\'e}sert},
  {Charbonneau}, {Batalha}, {Fortney}, {Rowe}, {Allen}, {Borucki}, {Brown},
  {Bryson}, {Ciardi}, {Cochran}, {Deming}, {Dunham}, {Fabrycky}, {Gautier},
  {Gilliland}, {Henze}, {Holman}, {Howell}, {Jenkins}, {Kinemuchi}, {Knutson},
  {Koch}, {Latham}, {Lissauer}, {Marcy}, {Ragozzine}, {Sasselov}, {Still},
  {Tenenbaum}, \& {Uddin}}]{Fressin11}
---. 2011, \apjs, 197, 5

\bibitem[{{Fressin} {et~al.}(2012){Fressin}, {Torres}, {Rowe}, {Charbonneau},
  {Rogers}, {Ballard}, {Batalha}, {Borucki}, {Bryson}, {Buchhave}, {Ciardi},
  {D{\'e}sert}, {Dressing}, {Fabrycky}, {Ford}, {Gautier}, {Henze}, {Holman},
  {Howard}, {Howell}, {Jenkins}, {Koch}, {Latham}, {Lissauer}, {Marcy},
  {Quinn}, {Ragozzine}, {Sasselov}, {Seager}, {Barclay}, {Mullally}, {Seader},
  {Still}, {Twicken}, {Thompson}, \& {Uddin}}]{Fressin12}
---. 2012, \nat, 482, 195

\bibitem[{{Gautier} {et~al.}(2012){Gautier}, {Charbonneau}, {Rowe}, {Marcy},
  {Isaacson}, {Torres}, {Fressin}, {Rogers}, {D{\'e}sert}, {Buchhave},
  {Latham}, {Quinn}, {Ciardi}, {Fabrycky}, {Ford}, {Gilliland}, {Walkowicz},
  {Bryson}, {Cochran}, {Endl}, {Fischer}, {Howell}, {Horch}, {Barclay},
  {Batalha}, {Borucki}, {Christiansen}, {Geary}, {Henze}, {Holman}, {Ibrahim},
  {Jenkins}, {Kinemuchi}, {Koch}, {Lissauer}, {Sanderfer}, {Sasselov},
  {Seager}, {Silverio}, {Smith}, {Still}, {Stumpe}, {Tenenbaum}, \& {Van
  Cleve}}]{Gautier12}
{Gautier}, III, T.~N., {et~al.} 2012, \apj, 749, 15

\bibitem[{{Gazak} {et~al.}(2012){Gazak}, {Johnson}, {Tonry}, {Dragomir},
  {Eastman}, {Mann}, \& {Agol}}]{Gazak12}
{Gazak}, J.~Z., {Johnson}, J.~A., {Tonry}, J., {Dragomir}, D., {Eastman}, J.,
  {Mann}, A.~W., \& {Agol}, E. 2012, Advances in Astronomy, 2012

\bibitem[{{Gilliland} {et~al.}(2010){Gilliland}, {Jenkins}, {Borucki},
  {Bryson}, {Caldwell}, {Clarke}, {Dotson}, {Haas}, {Hall}, {Klaus}, {Koch},
  {McCauliff}, {Quintana}, {Twicken}, \& {van Cleve}}]{Gilliland10}
{Gilliland}, R.~L., {et~al.} 2010, \apjl, 713, L160

\bibitem[{{Gilliland} {et~al.}(2013){Gilliland}, {Marcy}, {Rowe}, {Rogers},
  {Torres}, {Fressin}, {Lopez}, {Buchhave}, {Christensen-Dalsgaard},
  {D{\'e}sert}, {Henze}, {Isaacson}, {Jenkins}, {Lissauer}, {Chaplin}, {Basu},
  {Metcalfe}, {Elsworth}, {Handberg}, {Hekker}, {Huber}, {Karoff}, {Kjeldsen},
  {Lund}, {Lundkvist}, {Miglio}, {Charbonneau}, {Ford}, {Fortney}, {Haas},
  {Howard}, {Howell}, {Ragozzine}, \& {Thompson}}]{Gilliland13}
---. 2013, \apj, 766, 40

\bibitem[{{Gilliland} {et~al.}(2011){Gilliland}, {McCullough}, {Nelan},
  {Brown}, {Charbonneau}, {Nutzman}, {Christensen-Dalsgaard}, \&
  {Kjeldsen}}]{Gilliland11}
{Gilliland}, R.~L., {McCullough}, P.~R., {Nelan}, E.~P., {Brown}, T.~M.,
  {Charbonneau}, D., {Nutzman}, P., {Christensen-Dalsgaard}, J., \& {Kjeldsen},
  H. 2011, \apj, 726, 2

\bibitem[{{Goldreich} \& {Soter}(1966)}]{Goldreich66}
{Goldreich}, P., \& {Soter}, S. 1966, Icarus, 5, 375

\bibitem[{{Grillmair} {et~al.}(2012){Grillmair}, {Carey}, {Stauffer}, {Fisher},
  {Olds}, {Ingalls}, {Krick}, {Glaccum}, {Laine}, {Lowrance}, \&
  {Surace}}]{Grillmair12}
{Grillmair}, C.~J., {et~al.} 2012, in Society of Photo-Optical Instrumentation
  Engineers (SPIE) Conference Series, Vol. 8448, Society of Photo-Optical
  Instrumentation Engineers (SPIE) Conference Series

\bibitem[{{Handberg} \& {Campante}(2011)}]{Handberg11}
{Handberg}, R., \& {Campante}, T.~L. 2011, \aap, 527, A56

\bibitem[{{Henning} {et~al.}(2009){Henning}, {O'Connell}, \&
  {Sasselov}}]{Henning09}
{Henning}, W.~G., {O'Connell}, R.~J., \& {Sasselov}, D.~D. 2009, \apj, 707,
  1000

\bibitem[{{H{\o}g} {et~al.}(2000){H{\o}g}, {Fabricius}, {Makarov}, {Urban},
  {Corbin}, {Wycoff}, {Bastian}, {Schwekendiek}, \& {Wicenec}}]{Hog00}
{H{\o}g}, E., {et~al.} 2000, \aap, 355, L27

\bibitem[{{H{\o}g} {et~al.}(1998){H{\o}g}, {Kuzmin}, {Bastian}, {Fabricius},
  {Kuimov}, {Lindegren}, {Makarov}, \& {Roeser}}]{Hog98}
{H{\o}g}, E., {Kuzmin}, A., {Bastian}, U., {Fabricius}, C., {Kuimov}, K.,
  {Lindegren}, L., {Makarov}, V.~V., \& {Roeser}, S. 1998, \aap, 335, L65

\bibitem[{{Howard} {et~al.}(2013){Howard}, {Sanchis-Ojeda}, {Marcy}, {Johnson},
  {Winn}, {Isaacson}, {Fischer}, {Fulton}, {Sinukoff}, \& {Fortney}}]{Howard13}
{Howard}, A.~W., {et~al.} 2013, \nat, 503, 381

\bibitem[{{Howell} {et~al.}(2012){Howell}, {Rowe}, {Bryson}, {Quinn}, {Marcy},
  {Isaacson}, {Ciardi}, {Chaplin}, {Metcalfe}, {Monteiro}, {Appourchaux},
  {Basu}, {Creevey}, {Gilliland}, {Quirion}, {Stello}, {Kjeldsen},
  {Christensen-Dalsgaard}, {Elsworth}, {Garc{\'{\i}}a}, {Houdek}, {Karoff},
  {Molenda-{\.Z}akowicz}, {Thompson}, {Verner}, {Torres}, {Fressin}, {Crepp},
  {Adams}, {Dupree}, {Sasselov}, {Dressing}, {Borucki}, {Koch}, {Lissauer},
  {Latham}, {Buchhave}, {Gautier}, {Everett}, {Horch}, {Batalha}, {Dunham},
  {Szkody}, {Silva}, {Mighell}, {Holberg}, {Ballot}, {Bedding}, {Bruntt},
  {Campante}, {Handberg}, {Hekker}, {Huber}, {Mathur}, {Mosser}, {R{\'e}gulo},
  {White}, {Christiansen}, {Middour}, {Haas}, {Hall}, {Jenkins}, {McCaulif},
  {Fanelli}, {Kulesa}, {McCarthy}, \& {Henze}}]{Howell12}
{Howell}, S.~B., {et~al.} 2012, \apj, 746, 123

\bibitem[{{Huber} {et~al.}(2013){Huber}, {Chaplin}, {Christensen-Dalsgaard},
  {Gilliland}, {Kjeldsen}, {Buchhave}, {Fischer}, {Lissauer}, {Rowe},
  {Sanchis-Ojeda}, {Basu}, {Handberg}, {Hekker}, {Howard}, {Isaacson},
  {Karoff}, {Latham}, {Lund}, {Lundkvist}, {Marcy}, {Miglio}, {Silva Aguirre},
  {Stello}, {Arentoft}, {Barclay}, {Bedding}, {Burke}, {Christiansen},
  {Elsworth}, {Haas}, {Kawaler}, {Metcalfe}, {Mullally}, \&
  {Thompson}}]{Huber13}
{Huber}, D., {et~al.} 2013, \apj, 767, 127

\bibitem[{{Iglesias} \& {Rogers}(1996)}]{Iglesias96}
{Iglesias}, C.~A., \& {Rogers}, F.~J. 1996, \apj, 464, 943

\bibitem[{{Imbriani} {et~al.}(2004){Imbriani}, {Costantini}, {Formicola},
  {Bemmerer}, {Bonetti}, {Broggini}, {Corvisiero}, {Cruz}, {F{\"u}l{\"o}p},
  {Gervino}, {Guglielmetti}, {Gustavino}, {Gy{\"u}rky}, {Jesus}, {Junker},
  {Lemut}, {Menegazzo}, {Prati}, {Roca}, {Rolfs}, {Romano}, {Rossi Alvarez},
  {Sch{\"u}mann}, {Somorjai}, {Straniero}, {Strieder}, {Terrasi},
  {Trautvetter}, {Vomiero}, \& {Zavatarelli}}]{Imbriani04}
{Imbriani}, G., {et~al.} 2004, \aap, 420, 625

\bibitem[{{Imbriani} {et~al.}(2005){Imbriani}, {Costantini}, {Formicola},
  {Vomiero}, {Angulo}, {Bemmerer}, {Bonetti}, {Broggini}, {Confortola},
  {Corvisiero}, {Cruz}, {Descouvemont}, {F{\"u}l{\"o}p}, {Gervino},
  {Guglielmetti}, {Gustavino}, {Gy{\"u}rky}, {Jesus}, {Junker}, {Klug},
  {Lemut}, {Menegazzo}, {Prati}, {Roca}, {Rolfs}, {Romano}, {Rossi-Alvarez},
  {Sch{\"u}mann}, {Sch{\"u}rmann}, {Somorjai}, {Straniero}, {Strieder},
  {Terrasi}, \& {Trautvetter}}]{Imbriani05}
---. 2005, European Physical Journal A, 25, 455

\bibitem[{{Ingalls} {et~al.}(2012){Ingalls}, {Krick}, {Carey}, {Laine},
  {Surace}, {Glaccum}, {Grillmair}, \& {Lowrance}}]{Ingalls12}
{Ingalls}, J.~G., {Krick}, J.~E., {Carey}, S.~J., {Laine}, S., {Surace}, J.~A.,
  {Glaccum}, W.~J., {Grillmair}, C.~C., \& {Lowrance}, P.~J. 2012, in Society
  of Photo-Optical Instrumentation Engineers (SPIE) Conference Series, Vol.
  8442, Society of Photo-Optical Instrumentation Engineers (SPIE) Conference
  Series

\bibitem[{{Jenkins} {et~al.}(2010){Jenkins}, {Caldwell}, {Chandrasekaran},
  {Twicken}, {Bryson}, {Quintana}, {Clarke}, {Li}, {Allen}, {Tenenbaum}, {Wu},
  {Klaus}, {Van Cleve}, {Dotson}, {Haas}, {Gilliland}, {Koch}, \&
  {Borucki}}]{Jenkins10}
{Jenkins}, J.~M., {et~al.} 2010, \apjl, 713, L120

\bibitem[{{Kippenhahn} {et~al.}(2013){Kippenhahn}, {Weigert}, \&
  {Weiss}}]{Kippenhahn13}
{Kippenhahn}, R., {Weigert}, A., \& {Weiss}, A. 2013, {Stellar Structure and
  Evolution} (Berlin: Springer)

\bibitem[{{Kipping} {et~al.}(2014){Kipping}, {Nesvorn{\'y}}, {Buchhave},
  {Hartman}, {Bakos}, \& {Schmitt}}]{Kipping14}
{Kipping}, D.~M., {Nesvorn{\'y}}, D., {Buchhave}, L.~A., {Hartman}, J.,
  {Bakos}, G.~{\'A}., \& {Schmitt}, A.~R. 2014, \apj, 784, 28

\bibitem[{{Kipping} {et~al.}(2013){Kipping}, {Spiegel}, \&
  {Sasselov}}]{Kipping13}
{Kipping}, D.~M., {Spiegel}, D.~S., \& {Sasselov}, D.~D. 2013, MNRAS, accepted
  (arXiv:1306:3221)

\bibitem[{{Kjeldsen} {et~al.}(2008){Kjeldsen}, {Bedding}, \&
  {Christensen-Dalsgaard}}]{Kjeldsen08}
{Kjeldsen}, H., {Bedding}, T.~R., \& {Christensen-Dalsgaard}, J. 2008, \apjl,
  683, L175

\bibitem[{{Knutson} {et~al.}(2008){Knutson}, {Charbonneau}, {Allen}, {Burrows},
  \& {Megeath}}]{Knutson08}
{Knutson}, H.~A., {Charbonneau}, D., {Allen}, L.~E., {Burrows}, A., \&
  {Megeath}, S.~T. 2008, \apj, 673, 526

\bibitem[{{Knutson} {et~al.}(2007){Knutson}, {Charbonneau}, {Allen}, {Fortney},
  {Agol}, {Cowan}, {Showman}, {Cooper}, \& {Megeath}}]{Knutson07}
{Knutson}, H.~A., {et~al.} 2007, \nat, 447, 183

\bibitem[{{Kunz} {et~al.}(2002){Kunz}, {Fey}, {Jaeger}, {Mayer}, {Hammer},
  {Staudt}, {Harissopulos}, \& {Paradellis}}]{Kunz02}
{Kunz}, R., {Fey}, M., {Jaeger}, M., {Mayer}, A., {Hammer}, J.~W., {Staudt},
  G., {Harissopulos}, S., \& {Paradellis}, T. 2002, \apj, 567, 643

\bibitem[{{Lissauer} {et~al.}(2014){Lissauer}, {Marcy}, {Bryson}, {Rowe},
  {Jontof-Hutter}, {Agol}, {Borucki}, {Carter}, {Ford}, {Gilliland}, {Kolbl},
  {Star}, {Steffen}, \& {Torres}}]{Lissauer14}
{Lissauer}, J.~J., {et~al.} 2014, \apj, 784, 44

\bibitem[{{Mandel} \& {Agol}(2002)}]{Mandel02}
{Mandel}, K., \& {Agol}, E. 2002, \apjl, 580, L171

\bibitem[{{Marcus} {et~al.}(2010){Marcus}, {Sasselov}, {Hernquist}, \&
  {Stewart}}]{Marcus10}
{Marcus}, R.~A., {Sasselov}, D., {Hernquist}, L., \& {Stewart}, S.~T. 2010,
  \apjl, 712, L73

\bibitem[{{Marcy} {et~al.}(2014){Marcy}, {Isaacson}, {Howard}, {Rowe},
  {Jenkins}, {Bryson}, {Latham}, {Howell}, {Gautier}, {Batalha}, {Rogers},
  {Ciardi}, {Fischer}, {Gilliland}, {Kjeldsen}, {Christensen-Dalsgaard},
  {Huber}, {Chaplin}, {Basu}, {Buchhave}, {Quinn}, {Borucki}, {Koch}, {Hunter},
  {Caldwell}, {van Cleve}, {Kolbl}, {Weiss}, {Petigura}, {Seager}, {Morton},
  {Johnson}, {Ballard}, {Burke}, {Cochran}, {Endl}, {MacQueen}, {Everett},
  {Lissauer}, {Ford}, {Torres}, {Fressin}, {Brown}, {Steffen}, {Charbonneau},
  {Basri}, {Sasselov}, {Winn}, {Sanchis-Ojeda}, {Christiansen}, {Adams},
  {Henze}, {Dupree}, {Fabrycky}, {Fortney}, {Tarter}, {Holman}, {Tenenbaum},
  {Shporer}, {Lucas}, {Welsh}, {Orosz}, {Bedding}, {Campante}, {Davies},
  {Elsworth}, {Handberg}, {Hekker}, {Karoff}, {Kawaler}, {Lund}, {Lundkvist},
  {Metcalfe}, {Miglio}, {Silva Aguirre}, {Stello}, {White}, {Boss}, {DeVore},
  {Gould}, {Prsa}, {Agol}, {Barclay}, {Coughlin}, {Brugamyer}, {Mullally},
  {Quintana}, {Still}, {Thompson}, {Morrison}, {Twicken}, {Desert}, {Carter},
  {Crepp}, {Hebrard}, {Santerne}, {Moutou}, {Sobeck}, {Hudgins}, {Haas},
  {Robertson}, {Lillo-Box}, \& {Barrado}}]{Marcy14}
{Marcy}, G.~W., {et~al.} 2014, ApJS, accepted

\bibitem[{{Metcalfe} {et~al.}(2009){Metcalfe}, {Creevey}, \&
  {Christensen-Dalsgaard}}]{Metcalfe09}
{Metcalfe}, T.~S., {Creevey}, O.~L., \& {Christensen-Dalsgaard}, J. 2009, \apj,
  699, 373

\bibitem[{{Metcalfe} {et~al.}(2014){Metcalfe}, {Creevey}, {Dogan}, {Mathur},
  {Xu}, {Bedding}, {Chaplin}, {Christensen-Dalsgaard}, {Karoff}, {Trampedach},
  {Benomar}, {Brown}, {Buzasi}, {Campante}, {Celik}, {Cunha}, {Davies},
  {Deheuvels}, {Derekas}, {Di Mauro}, {Garcia}, {Guzik}, {Howe}, {MacGregor},
  {Mazumdar}, {Montalban}, {Monteiro}, {Salabert}, {Serenelli}, {Stello},
  {Steslicki}, {Suran}, {Yildiz}, {Aksoy}, {Elsworth}, {Gruberbauer},
  {Guenther}, {Lebreton}, {Molaverdikhani}, {Pricopi}, {Simoniello}, \&
  {White}}]{Metcalfe14}
{Metcalfe}, T.~S., {et~al.} 2014, submitted to ApJ (arXiv:1402.3614)

\bibitem[{{Mordasini} {et~al.}(2012){Mordasini}, {Alibert}, {Georgy},
  {Dittkrist}, {Klahr}, \& {Henning}}]{Mordasini12}
{Mordasini}, C., {Alibert}, Y., {Georgy}, C., {Dittkrist}, K.-M., {Klahr}, H.,
  \& {Henning}, T. 2012, \aap, 547, A112

\bibitem[{{Morgan} \& {Anders}(1980)}]{Morgan80}
{Morgan}, J.~W., \& {Anders}, E. 1980, Proceedings of the National Academy of
  Science, 77, 6973

\bibitem[{{Morton}(2012)}]{Morton12}
{Morton}, T.~D. 2012, \apj, 761, 6

\bibitem[{{Morton} \& {Johnson}(2011)}]{Morton11}
{Morton}, T.~D., \& {Johnson}, J.~A. 2011, \apj, 738, 170

\bibitem[{{Nutzman} {et~al.}(2011){Nutzman}, {Gilliland}, {McCullough},
  {Charbonneau}, {Christensen-Dalsgaard}, {Kjeldsen}, {Nelan}, {Brown}, \&
  {Holman}}]{Nutzman11}
{Nutzman}, P., {et~al.} 2011, \apj, 726, 3

\bibitem[{{Paxton} {et~al.}(2011){Paxton}, {Bildsten}, {Dotter}, {Herwig},
  {Lesaffre}, \& {Timmes}}]{Paxton11}
{Paxton}, B., {Bildsten}, L., {Dotter}, A., {Herwig}, F., {Lesaffre}, P., \&
  {Timmes}, F. 2011, \apjs, 192, 3

\bibitem[{{Paxton} {et~al.}(2013){Paxton}, {Cantiello}, {Arras}, {Bildsten},
  {Brown}, {Dotter}, {Mankovich}, {Montgomery}, {Stello}, {Timmes}, \&
  {Townsend}}]{Paxton13}
{Paxton}, B., {et~al.} 2013, ApJS, accepted (arXiv:1301.0319)

\bibitem[{{Pepe} {et~al.}(2013){Pepe}, {Cameron}, {Latham}, {Molinari}, {Udry},
  {Bonomo}, {Buchhave}, {Charbonneau}, {Cosentino}, {Dressing}, {Dumusque},
  {Figueira}, {Fiorenzano}, {Gettel}, {Harutyunyan}, {Haywood}, {Horne},
  {Lopez-Morales}, {Lovis}, {Malavolta}, {Mayor}, {Micela}, {Motalebi},
  {Nascimbeni}, {Phillips}, {Piotto}, {Pollacco}, {Queloz}, {Rice}, {Sasselov},
  {S{\'e}gransan}, {Sozzetti}, {Szentgyorgyi}, \& {Watson}}]{Pepe13}
{Pepe}, F., {et~al.} 2013, \nat, 503, 377

\bibitem[{{Petigura} {et~al.}(2013{\natexlab{a}}){Petigura}, {Howard}, \&
  {Marcy}}]{Petigura13a}
{Petigura}, E.~A., {Howard}, A.~W., \& {Marcy}, G.~W. 2013{\natexlab{a}},
  Proceedings of the National Academy of Science, 110, 19273

\bibitem[{{Petigura} {et~al.}(2013{\natexlab{b}}){Petigura}, {Marcy}, \&
  {Howard}}]{Petigura13b}
{Petigura}, E.~A., {Marcy}, G.~W., \& {Howard}, A.~W. 2013{\natexlab{b}}, \apj,
  770, 69

\bibitem[{{Rogers} \& {Nayfonov}(2002)}]{Rogers02}
{Rogers}, F.~J., \& {Nayfonov}, A. 2002, \apj, 576, 1064

\bibitem[{{Rogers}(2014)}]{Rogers14}
{Rogers}, L.~A. 2014, submitted to ApJ

\bibitem[{{Rowe} {et~al.}(2014){Rowe}, {Bryson}, {Marcy}, {Lissauer},
  {Jontof-Hutter}, {Mullally}, {Gilliland}, {Issacson}, {Ford}, {Howell},
  {Borucki}, {Haas}, {Huber}, {Steffen}, {Thompson}, {Quintana}, {Barclay},
  {Still}, {Fortney}, {Gautier}, {Hunter}, {Caldwell}, {Ciardi}, {Devore},
  {Cochran}, {Jenkins}, {Agol}, {Carter}, \& {Geary}}]{Rowe14}
{Rowe}, J.~F., {et~al.} 2014, \apj, 784, 45

\bibitem[{{Roxburgh} \& {Vorontsov}(2003)}]{Roxburgh03}
{Roxburgh}, I.~W., \& {Vorontsov}, S.~V. 2003, \aap, 411, 215

\bibitem[{{Sanchis-Ojeda} {et~al.}(2013){Sanchis-Ojeda}, {Rappaport}, {Winn},
  {Levine}, {Kotson}, {Latham}, \& {Buchhave}}]{SanchisOjeda13}
{Sanchis-Ojeda}, R., {Rappaport}, S., {Winn}, J.~N., {Levine}, A., {Kotson},
  M.~C., {Latham}, D.~W., \& {Buchhave}, L.~A. 2013, \apj, 774, 54

\bibitem[{{Seager}(2010)}]{Seager10}
{Seager}, S. 2010, {Exoplanet Atmospheres: Physical Processes}

\bibitem[{{Seager} \& {Mall{\'e}n-Ornelas}(2003)}]{Seager03}
{Seager}, S., \& {Mall{\'e}n-Ornelas}, G. 2003, \apj, 585, 1038

\bibitem[{{Silva Aguirre} {et~al.}(2011){Silva Aguirre}, {Ballot}, {Serenelli},
  \& {Weiss}}]{Silva11}
{Silva Aguirre}, V., {Ballot}, J., {Serenelli}, A.~M., \& {Weiss}, A. 2011,
  \aap, 529, A63

\bibitem[{{Silva Aguirre} {et~al.}(2013){Silva Aguirre}, {Basu}, {Brand{\~a}o},
  {Christensen-Dalsgaard}, {Deheuvels}, {Do{\u g}an}, {Metcalfe}, {Serenelli},
  {Ballot}, {Chaplin}, {Cunha}, {Weiss}, {Appourchaux}, {Casagrande},
  {Cassisi}, {Creevey}, {Garc{\'{\i}}a}, {Lebreton}, {Noels}, {Sousa},
  {Stello}, {White}, {Kawaler}, \& {Kjeldsen}}]{Silva13}
{Silva Aguirre}, V., {et~al.} 2013, \apj, 769, 141

\bibitem[{{Smith} {et~al.}(2012){Smith}, {Stumpe}, {Van Cleve}, {Jenkins},
  {Barclay}, {Fanelli}, {Girouard}, {Kolodziejczak}, {McCauliff}, {Morris}, \&
  {Twicken}}]{Smith12}
{Smith}, J.~C., {et~al.} 2012, \pasp, 124, 1000

\bibitem[{{Sozzetti} {et~al.}(2007){Sozzetti}, {Torres}, {Charbonneau},
  {Latham}, {Holman}, {Winn}, {Laird}, \& {O'Donovan}}]{Sozzetti07}
{Sozzetti}, A., {Torres}, G., {Charbonneau}, D., {Latham}, D.~W., {Holman},
  M.~J., {Winn}, J.~N., {Laird}, J.~B., \& {O'Donovan}, F.~T. 2007, \apj, 664,
  1190

\bibitem[{{Steffen} {et~al.}(2012){Steffen}, {Fabrycky}, {Ford}, {Carter},
  {D{\'e}sert}, {Fressin}, {Holman}, {Lissauer}, {Moorhead}, {Rowe},
  {Ragozzine}, {Welsh}, {Batalha}, {Borucki}, {Buchhave}, {Bryson}, {Caldwell},
  {Charbonneau}, {Ciardi}, {Cochran}, {Endl}, {Everett}, {Gautier},
  {Gilliland}, {Girouard}, {Jenkins}, {Horch}, {Howell}, {Isaacson}, {Klaus},
  {Koch}, {Latham}, {Li}, {Lucas}, {MacQueen}, {Marcy}, {McCauliff}, {Middour},
  {Morris}, {Mullally}, {Quinn}, {Quintana}, {Shporer}, {Still}, {Tenenbaum},
  {Thompson}, {Twicken}, \& {van Cleve}}]{Steffen12}
{Steffen}, J.~H., {et~al.} 2012, \mnras, 2482

\bibitem[{{Stevenson} {et~al.}(2012){Stevenson}, {Harrington}, {Fortney},
  {Loredo}, {Hardy}, {Nymeyer}, {Bowman}, {Cubillos}, {Bowman}, \&
  {Hardin}}]{Stevenson12}
{Stevenson}, K.~B., {et~al.} 2012, \apj, 754, 136

\bibitem[{{Stumpe} {et~al.}(2012){Stumpe}, {Smith}, {Van Cleve}, {Twicken},
  {Barclay}, {Fanelli}, {Girouard}, {Jenkins}, {Kolodziejczak}, {McCauliff}, \&
  {Morris}}]{Stumpe12}
{Stumpe}, M.~C., {et~al.} 2012, \pasp, 124, 985

\bibitem[{{Swift} {et~al.}(2013){Swift}, {Johnson}, {Morton}, {Crepp},
  {Montet}, {Fabrycky}, \& {Muirhead}}]{Swift13}
{Swift}, J.~J., {Johnson}, J.~A., {Morton}, T.~D., {Crepp}, J.~R., {Montet},
  B.~T., {Fabrycky}, D.~C., \& {Muirhead}, P.~S. 2013, \apj, 764, 105

\bibitem[{{Tegmark} {et~al.}(2004){Tegmark}, {Strauss}, {Blanton}, {Abazajian},
  {Dodelson}, {Sandvik}, {Wang}, {Weinberg}, {Zehavi}, {Bahcall}, {Hoyle},
  {Schlegel}, {Scoccimarro}, {Vogeley}, {Berlind}, {Budavari}, {Connolly},
  {Eisenstein}, {Finkbeiner}, {Frieman}, {Gunn}, {Hui}, {Jain}, {Johnston},
  {Kent}, {Lin}, {Nakajima}, {Nichol}, {Ostriker}, {Pope}, {Scranton},
  {Seljak}, {Sheth}, {Stebbins}, {Szalay}, {Szapudi}, {Xu}, {Annis},
  {Brinkmann}, {Burles}, {Castander}, {Csabai}, {Loveday}, {Doi}, {Fukugita},
  {Gillespie}, {Hennessy}, {Hogg}, {Ivezi{\'c}}, {Knapp}, {Lamb}, {Lee},
  {Lupton}, {McKay}, {Kunszt}, {Munn}, {O'Connell}, {Peoples}, {Pier},
  {Richmond}, {Rockosi}, {Schneider}, {Stoughton}, {Tucker}, {vanden Berk},
  {Yanny}, \& {York}}]{Tegmark04}
{Tegmark}, M., {et~al.} 2004, \prd, 69, 103501

\bibitem[{{Thoul} {et~al.}(1994){Thoul}, {Bahcall}, \& {Loeb}}]{Thoul94}
{Thoul}, A.~A., {Bahcall}, J.~N., \& {Loeb}, A. 1994, \apj, 421, 828

\bibitem[{{Torres} {et~al.}(2012){Torres}, {Fischer}, {Sozzetti}, {Buchhave},
  {Winn}, {Holman}, \& {Carter}}]{Torres12}
{Torres}, G., {Fischer}, D.~A., {Sozzetti}, A., {Buchhave}, L.~A., {Winn},
  J.~N., {Holman}, M.~J., \& {Carter}, J.~A. 2012, \apj, 757, 161

\bibitem[{{Torres} {et~al.}(2004){Torres}, {Konacki}, {Sasselov}, \&
  {Jha}}]{Torres04}
{Torres}, G., {Konacki}, M., {Sasselov}, D.~D., \& {Jha}, S. 2004, \apj, 614,
  979

\bibitem[{{Torres} {et~al.}(2008){Torres}, {Winn}, \& {Holman}}]{Torres08}
{Torres}, G., {Winn}, J.~N., \& {Holman}, M.~J. 2008, \apj, 677, 1324

\bibitem[{{Van Eylen} {et~al.}(2014){Van Eylen}, {Lund}, {Silva Aguirre},
  {Arentoft}, {Kjeldsen}, {Albrecht}, {Chaplin}, {Isaacson}, {Pedersen},
  {Jessen-Hansen}, {Tingley}, {Christensen-Dalsgaard}, {Aerts}, {Campante}, \&
  {Bryson}}]{Vaneylen14}
{Van Eylen}, V., {et~al.} 2014, \apj, 782, 14

\bibitem[{{Weiss} \& {Schlattl}(2008)}]{Weiss08}
{Weiss}, A., \& {Schlattl}, H. 2008, \apss, 316, 99

\bibitem[{{Winn} {et~al.}(2009){Winn}, {Johnson}, {Albrecht}, {Howard},
  {Marcy}, {Crossfield}, \& {Holman}}]{Winn09}
{Winn}, J.~N., {Johnson}, J.~A., {Albrecht}, S., {Howard}, A.~W., {Marcy},
  G.~W., {Crossfield}, I.~J., \& {Holman}, M.~J. 2009, \apjl, 703, L99

\bibitem[{{Woitaszek} {et~al.}(2009){Woitaszek}, {Metcalfe}, \&
  {Shorrock}}]{Woitaszek09}
{Woitaszek}, M., {Metcalfe}, T., \& {Shorrock}, I. 2009, in Proceedings of the
  5th Grid Computing Environments Workshop (New York: ACM), p. 1-7,
  doi:10.1145/1658260.1658262

\bibitem[{{Yoder}(1995)}]{Yoder95}
{Yoder}, C.~F. 1995, in Global Earth Physics: A Handbook of Physical Constants,
  ed. T.~J. {Ahrens}, 1

\bibitem[{{Zeng} \& {Sasselov}(2013)}]{Zeng13}
{Zeng}, L., \& {Sasselov}, D. 2013, \pasp, 125, 227

\end{thebibliography}

\clearpage
\newpage

%\begin{figure}[h!]
%\begin{center}
% \includegraphics[width=5in]{koi069_mcmc_newrho_tap_ldc.eps} 
% \caption{Quadratic limb darkening coefficients inferred for Kepler-93,
 %  derived from the fit to the transit light curve. The grey contours
  % depict the 1 and 2$\sigma$ confidence intervals inferred from
  % fitting the transit with no prior on the stellar density. The red
  % contour depicts the 1$\sigma$ confidence interval inferred after
 %  applying the asteroseismic prior. In green and blue, we show the theoretical limb
 %  darkening values from \cite{Claret13} for a star similar to Kepler-93b.}
%  \label{fig:ldc}
%\end{center}
%\end{figure}

%\begin{figure}[h!]
%\begin{center}
 %\includegraphics[width=5in]{kipping_mah_koi069.eps} 
 %\caption{{\it Left panel:} Posterior distribution of the quantity
 %  $P_{MAH}$ as defined by \cite{Kipping13}, which is physically
 %  attributable to the extent of an exoplanet's radius behond the
 %  minimum density core comprised of water. With this methodology, the
 %  measured mass and radius of the planet imply that 
 %it has a 3\% chance of possessing an atmosphere. {\it Right panels:} The
 %  posteriors of mass and radius for Kepler-93b that we employed to
 %  produce the posterior at left.}
 % \label{fig:kipping_mah}
%\end{center}
%\end{figure}

%\end{document}

%\end{thebibliography}

\appendix

\section{Estimation of stellar properties using oscillation frequencies}
\label{sec:est}

Five members of the team (S.B., J.C.-D., T.S.M., V.S.A and D.S.)
performed a detailed modeling of the host star using estimates of the
individual oscillation frequencies and spectroscopic parameters as
input. The estimated properties from the global asteroseismic
parameters analysis presented in \cite{Huber13} were used either
as starting guesses or as a guideline check for initial results.

S.B. and J.C.-D. adopted the same analysis procedures discussed in
detail in \cite{Chaplin13}, using, respectively, models from the
Yale stellar evolution code, YREC \citep{Demarque08} and the ASTEC
code \citep{Christensen08a} (for J.C.-D.'s analysis, see also
\citealt{Silva13}). T.S.M. used the Asteroseismic Modeling
Portal (AMP) \citep{Metcalfe09, Woitaszek09}, a
web-based tool linked to TeraGrid computing resources that runs an
automated search based on a parallel genetic algorithm. Further
details on the application to asteroseismic data may again be found in
\cite{Chaplin13}. Here, in addition to the individual oscillation
frequencies, two sets of frequency ratios---$r_{02}(n)$ and
$r_{010}(n)$ \citep{Roxburgh03}---were also used as seismic
inputs (see also \citealt{Metcalfe14} for further information).

V.S.A. computed models using the Garching Stellar Evolution Code
(GARSTEC, \citealt{Weiss08}). The input physics included the OPAL
equation of state \citep{Rogers02}, OPAL opacities at
high temperatures \citep{Iglesias96} and \cite{Ferguson05}
opacities at low temperatures. The nuclear reaction rates were those
of the NACRE compilation \citep{Angulo99}, including the updated
cross section for $^{14}\mathrm{N}(p,\gamma)^{15}\mathrm{O}$ from
\cite{Formicola04}. Convective regions were treated under the
mixing-length formalism of \cite{Kippenhahn13}, while diffusion
of helium and heavy elements was included following the prescription
of \cite{Thoul94}.

A grid of stellar models was constructed spanning a mass range of 0.8
to $1.0\,\rm M_{\odot}$, with a varying initial helium abundance
between 0.22 and 0.30. Five different compositions were determined for
each helium value, which covered the 1-$\sigma$ range in metallicity
determined from the spectroscopy, at the observed average large
frequency separation. The best-matching model was determined from a
fit to the spectroscopic data and the frequency separation ratios
$r_{010}$ and $r_{02}$ (e.g., \citealt{Roxburgh03}, \citealt{Silva11}). Final properties and uncertainties were
determined from weighted means and standard deviations of properties
of all models in the selected grid, with weights fixed by the $\chi^2$
of the separation ratios and spectroscopic values (with allowance made
for the correlations between the separation ratios; see \citealt{Silva13} for a detailed description).

D.S. made use of the MESA stellar evolution code \citep{Paxton11}. The ``astero'' mode of MESA has several ways of matching
stellar models to observational data, including data from spectroscopy
and asteroseismology \citep{Paxton13}.  The method used in this
work employs an automated simplex search algorithm to optimize the
model parameters that best reproduce the observations. Each model
evaluation was initiated in the fully convective, pre-main-sequence
stage, and evolved until the $\chi^2$ between the model parameters and
observational parameters had become significantly larger than its
minimum value along that particular evolutionary track.  Repeated
tracks with different initial values for mass, helium content, [Fe/H],
and mixing length parameter were controlled by the simplex algorithm,
based on the minimum $\chi^2$ of previous tracks.  The $\chi^2$
comprised contributions from the asteroseismic data (frequencies and
the frequency ratios, $r_{02}$ and $r_{010}$) and from spectroscopy
($T_\mathrm{eff}$, [Fe/H]).  Model frequencies were calculated by
ADIPLS \citep{Christensen08b} from an internal call within
MESA, and were corrected for near-surface effects according to the
prescription by \cite{Kjeldsen08}.  The uncertainties of the
best-fitting model properties were found by evaluating locations on the
$\chi^2$ surface where min($\chi^2+1$) was reached.

We used the OPAL equation of state \citep{Rogers02}, the OPAL
opacities \citep{Iglesias96}, with low-temperature opacities
from \cite{Ferguson05}.  Nuclear reaction rates were from NACRE
\citep{Angulo99}, with updates for $^{14}{\rm
N}(p,\gamma)^{15}{\rm O}$ \citep{Imbriani04, Imbriani05} and $^{12}{\rm
C}(\alpha,\gamma)^{16}{\rm O}$ \citep{Kunz02}.  Models used a
fixed value for exponential overshoot ($f_{\mathrm{ov}}$ = 0.015,
roughly equivalent to a linear (fully mixed) overshoot of 0.15 times
the pressure scale height.  Diffusion was not turned on.  Convection
was treated according to mixing-length theory. We refer the reader to
\cite{Paxton11} for further details.

\end{document}